\documentclass[preprint,floatfix,showpacs]{revtex4}
\usepackage{color}

\usepackage{mathbbol}              % for math symbols.
\usepackage{graphics,graphicx,epsf,ulem,amsmath,eucal,bm}

\def\kket#1{\mathinner{|{#1}\rangle\rangle}}

\newcommand{\rrangle}{\rangle\rangle}

\begin{document}

\title{Sidebands shifts and induced sidebands in rf-dressed Rydberg systems}
%\date{\today}
\author{M. Tanasittikosol}
\altaffiliation[Present address:]{Theoretical and Computational Physics (TCP)
Group, Department of Physics,
King Mongkut's University of Technology Thonburi, Bangkok 10140, Thailand.}
\author{R. M. Potvliege} 
\email{r.m.potvliege@durham.ac.uk}
\affiliation{Joint Quantum Centre (JQC) Durham-Newcastle,
 Department of Physics, Durham University,
South Road, Durham DH1 3LE, UK}

\begin{abstract}
The effect of an ac modulation on a 2-level or 3-level
system is studied theoretically.
The absorption spectrum is calculated by
solving the optical Bloch equations and
is interpreted by reference to the Floquet quasienergy spectrum. 
The dependence of the absorption sidebands on the intensity
of the coupling laser field in Rydberg systems
submitted to a radio-frequency (rf) field is analysed in detail.
It is shown that for sufficiently strong coupling fields
additional sidebands appear in the probe absorption spectrum in ladder
3-level systems. These additional sidebands are
induced by the coupling
of the intermediate state to the Floquet manifold spawned by
the upper state under rf modulation.
\end{abstract}

\maketitle

%%%%%%%%%%%%%%%%%%%%%%%%%%%%%%%%%%%%%%%%%%%%%%%%%%%%%%%%%%%%%%%%%%%%%%%%%%%%%%%%%%%%%%%%%%%%%%%%%%%%%%%%%%%%%%%%%%%%%%%%%%%%%%%%%%%%%%%%%%%%%%%%%%%%%%%%%%%%%%%%%%%%%%%%%%%%%%%%%%%%%%%%%%%%%%%%%%%%%%%%%%%%%%%%%%%%%%%%%%%%%%%%%%%%%%%%%%%%%%%%%%%%%%%%%%%%%%%%%%%%%%%%%%%%%%%%%%%%%%%%%%%%%%%%%%%%%%%%%%%%%%%%
%%%%%%%%%%%%%%%%%%%%%%%%%%%%%%%%%%%%%%%%%%%%%%%%%%%%%%%%%%%%%%%%%%%%%%%%%%%%%%%%%%%%%%%%%%%%%%%%%%%%%%%%%%%%%%%%%%%%%%%%%%%%%%%%%%%%%%%%%%%%%%%%%%%%%%%%%%%%%%%%%%%%%%%%%%%%%%%%%%%%%%%%%%%%%%%%%%%%%%%%%%%%%%%%%%%%%%%%%%%%%%%%%%%%%%%%%%%%%%%%%%%%%%%%%%%%%%%%%%%%%%%%%%%%%%%%%%%%%%%%%%%%%%%%%%%%%%%%%%%%%%%%
\section{Introduction}\label{sec:1}
There is currently much interest for electromagnetically induced transparency
(EIT)
in Rydberg atoms \cite {eit_review}, in particular for the 
cooperative atom-light and photon-photon dynamics found in Rydberg systems 
\cite{polariton,pritchard10,gorshkov11,petrosyan11} and for their possible
application, e.g., in 
quantum information processing 
\cite{saffman}, in the production of 
correlated photon pairs \cite{pritchard12}, and in metrology 
\cite{osterwalder,mohapatra}.
The present work is motivated by a recent experimental study
of EIT in a 3-level Rydberg system with 
radio frequency (rf) modulation \cite{bason}.
%%Much of the recent interest for
%%EIT \cite{eit_review} in Rydberg systems stems from the associated
%%cooperative atom-light and photon-photon interactions it
%%\cite{polariton,pritchard10,gorshkov11,petrosyan11}. Besides their
%%intrinsic interest, these systems have potential applications in 
%%quantum information processing in terms of photonic quantum
%%gates and quantum memories \cite{saffman} and the production of
%%correlated photon pairs \cite{pritchard12}.
In this experiment, 
the absorption on the $5S_{1/2}$--$5P_{3/2}$
transition of ${}^{87}$Rb was probed by a weak laser field
in the presence of both an rf field and a strong laser field resonant
with the transition between the intermediate
$5P_{3/2}(F'=3)$ state and the highly polarizable
$32S_{1/2}(F'=2)$ state. The rf field was too low in frequency for
driving multiphoton transitions between the lower, intermediate
or upper states and other bound states. However, it was
strong enough to Stark-shift the upper state periodically by tens of MHz,
which resulted into the EIT features in the probe
absorption spectrum shifting and splitting into multiple EIT sidebands.

The existence of absorption sidebands is a well known feature
of ac-modulated 2-level systems \cite{budker}, including 2-level 
atomic or molecular systems 
\cite{blochinzew,
townes,autler,bayf81,linden,zhang94,takeshi,ditz09}.
The effect of an oscillating electric field on molecular absorption lines
with quadratic Stark shift was described by Townes and Merritt
in 1947 \cite{townes}: Lines whose frequency width is much larger
than the 
frequency of modulation 
adiabatically follow the oscillations of the field --- i.e., a state
of dipole polarizability $\alpha$ in a non-resonant ac field
${\cal E}(t)= {\cal E}_{0}\sin(\omega_0 t)$
is simply displaced in energy by
$\alpha [{\cal E}(t)]^2/2$ at time $t$.
However, lines whose frequency width is of the same order or smaller than
the modulation frequency split
into manifolds of components displaced in energy from their zero-field
limit by
$\alpha {\cal E}_{0}^2/4+2n\hbar\omega_0$, $n=0,\pm 1, \pm 2,\ldots$, and the
intensity of these components varies with $n$, $\omega_0$ and ${\cal E}_0$
like $J_n^2(\alpha {\cal E}_{0}/8\hbar \omega_0)$.
Analogous shifts and sidebands are also found in systems
governed by an Hamiltonian unitarily equivalent to that governing ac-modulated
2-level atomic or molecular systems,
such as ac-driven superconducting flux qubits
\cite{Vion,chu0,Chiorescu,Collin,Oliver,chu,Tuorila,hausinger}.

Interactions with other fields will
in general introduce additional
$n$-dependent Stark shifts. For this reason, the separation between adjacent
sidebands of an ac modulated 2-level system
is not exactly $2\hbar\omega_0$ and the intensities of the absorption
lines do not exactly follow a $J_n^2(\alpha {\cal E}_{0}/8\hbar \omega_0)$
law. The correct separations and scalings
can be obtained by combining Floquet
theory and Van Vleck perturbation theory
\cite{chu,hausinger}. However,
these deviations from the results of Townes and Merritt are significant only
when the interaction coupling the two states of the system is sufficiently
strong. Perhaps for this reason, they do not appear to have been previously
considered
in an atomic and molecular physics context \cite{bason2}.

In many ac-modulated 2-level atomic systems of practical interest,
one of the two states is much more polarizable than the other and
the frequency widths of both the state and the coupling laser
field are small compared 
to the modulation frequency. In these conditions, it
may be tempting to
ignore the manifold structure of the less polarizable state, retain
that of the more polarizable state, and model
the ac-modulated two-level system as a 
many-level system in which
the less polarizable state interacts with a manifold of independent
sideband states. We refer to this approximate description 
as the $N$-level approximation. It
leads to the results described by Townes and Merritt \cite{townes}
in the limit of a weak interaction between the states and
makes it possible to calculate the absorption spectrum rapidly at a minimal
cost. However,
as is shown below, this approximation breaks down for strong coupling
between the two states: 
In order to obtain the correct results, the sideband structure
of {\em both} states must be properly taken into account. This can be done,
e.g., by carrying out a full Floquet analysis
of the ac-modulated 2-level system
\cite{chu,hausinger,shirley,barone,book,incomm}, or by solving the optical Bloch
equations with time-dependent modulation. The sidebands of the less
polarizable state here originate from the coupling of this state to the
Floquet manifold formed by the
other state, not from a direct effect of the ac-modulation. Nonetheless,
as is illustrated below, they may manifest by well defined, additional resonance
structures in absorption spectra. We refer to these as induced sidebands.

In this paper, we investigate the changes in the sideband structure of
the absorption spectrum arising from the interaction with the coupling
laser field, both in two-level and three-level systems. The absorption spectrum
is calculated by solving the optical Bloch equations and is interpreted 
by reference to the Floquet quasienergy spectrum. The $N$-level approximation
is compared to the exact results throughout the paper.
For simplicity, we first focus
on ac-modulated two-level systems
(Sec.\ \ref{sec:3}). The coupling with
a lower state is then introduced. The probe absorption in an rf-modulated
3-state 
ladder system and the appearance of induced sidebands
is studied in Sec.\ \ref{sec:4}. In particular,
we give examples of 
sidebands shift in EIT spectra for both cold and thermal ensembles.
We conclude with a summary of our main findings (Sec.\ \ref{sec:5}).

We use atomic units throughout the rest of this paper, except where indicated
otherwise.
%%%%%%%%%%%%%%%%%%%%%%%%%%%%%%%%%%%%%%%%%%%%%%%%%%%%%%%%%%%%%%%%%%%%%%%%%%%%%%%%%%%%%%%%%%%%%%%%%%%%%%%%%%%%%%%%%%%%%%%%%%%%%%%%%%%%%%%%%%%%%%%%%%%%%%%%%%%%%%%%%%%%%%%%%%%%%%%%%%%%%%%%%%%%%%%%%%%%%%%%%%%%%%%%%%%%%%%%%%%%%%%%%%%%%%%%%%%%%%%%%%%%%%%%%%%%%%%%%%%%%%%%%%%%%%%%%%%%%%%%%%%%%%%%%%%%%%%%%%%%%%%%
\section{RF-dressed two-level systems}\label{sec:3}
\subsection{RF-dressing in the adiabatic approximation}\label{subsec:3a0}

We first concentrate on the rf dynamics of the subsystem
formed by the intermediate and upper states of our 3-level ladder Rydberg
system, states $|b\rangle$ and $|c\rangle$.
We thus assume, for the time being,
that neither $|b\rangle$ nor $|c\rangle$ interact with the lower state of the
system, $|a\rangle$. 
Making the rotating wave approximation for the laser field and
passing to slowly-varying
variables would then reduce the time-dependent Schr\"{o}dinger equation to
\begin{equation}
\frac{{d}\;}{{d}t}\begin{pmatrix} c_{b}\\c_{c}\end{pmatrix}=\begin{pmatrix}0& -{i}\Omega_{\text{c}}/2\\ -{i}\Omega_{\text{c}}/2& {i}
\Delta_{\text{c}}\end{pmatrix}\begin{pmatrix}c_{b}\\c_{c}\end{pmatrix}\label{eq:HT0}
\end{equation}
%%%%%%%%%%%%%%%%%
%%%%%%%%%%%%%%%%%
if the rf field was absent. $\Omega_{\rm c}$ and $\Delta_{\rm c}$
are respectively the Rabi frequency \cite{Rabi} (which we assume real)
and the detuning of
the control laser field coupling the states $|b\rangle$ and $|c\rangle$
in the ladder system: Denoting the zero-field energies of these two states
by $w_b$ and $w_c$ and the angular frequency of the control laser by
$\omega_{\rm c}$,  $\Delta_{\rm c}=\omega_{\rm c}-\omega_{cb}$ with
$\omega_{cb}=w_c-w_b$. The state vector of the system,
$|\Psi(t)\rangle$,
is a linear superposition
of the bare states $|b\rangle$ and $|c\rangle$
and
the functions $c_b(t)$ and $c_c(t)$ are the respective probability amplitudes:
\begin{equation}
|\Psi(t)\rangle = c_b(t)|b\rangle + c_c(t)|c\rangle.
\end{equation}

We write the electric field component of the rf field as
\begin{equation}
{\cal E}(t)= {\cal E}_{\rm rf}\sin(\omega_{\rm rf} t).
\end{equation}
We assume that the angular frequency $\omega_{\rm rf}$
is so much smaller than $\omega_{bc}$, and so much smaller than the
transition frequencies between $|b\rangle$ and $|c\rangle$
and any other unperturbed state, that $|b\rangle$ and $|c\rangle$
evolve adiabatically
under the effect of this field.
Hence, we replace
the time-dependent bare states $\exp(-{\rm i}w_b t)|b\rangle$ and
$\exp(-{\rm i}w_c t)|c\rangle$ by the adiabatic states
\cite{budker,blochinzew,autler,Pont}
\begin{equation}
|B(t)\rangle = \exp\left(-{i} \int^t E_b[{\cal E}(t')]\,{d}t'
\right) |b[{\cal E}(t)]\rangle
\end{equation}
and
\begin{equation}
|C(t)\rangle = \exp\left(-{i} \int^t E_c[{\cal E}(t')]\,{d}t'
\right) |c[{\cal E}(t)]\rangle.
\end{equation}
In these expressions, $|b[{\cal E}]\rangle$ and 
$|c[{\cal E}]\rangle$ denote the state vectors which develop adiabatically
from the unperturbed states $|b\rangle$ and $|c\rangle$ when a static electric
field is turned on fromn 0 to ${\cal E}$, and 
$E_b[{\cal E}]$ and $E_c[{\cal E}]$ are the corresponding eigenenergies
of the Stark Hamiltonian. The rf field is assumed to be weak enough that,
in sufficiently good approximation,
$E_b[{\cal E}(t)]=w_b-\alpha_b [{\cal E}(t)]^2/2$ and
$E_c[{\cal E}(t)]=w_c-\alpha_c [{\cal E}(t)]^2/2$, and moreover that
$|b[{\cal E}(t)]\rangle=
|b\rangle+{\cal E}(t)|b^{(1)}\rangle$ and
$|c[{\cal E}(t)]\rangle=
|c\rangle+{\cal E}(t)|c^{(1)}\rangle$. The two state vectors
$|b^{(1)}\rangle$ and $|c^{(1)}\rangle$ are defined by first order perturbation
theory, and $\alpha_b$ and $\alpha_c$ are the static dipole polarizability of
the states $|b\rangle$ and $|c\rangle$.
For the applications we have in mind, $|\alpha_c| \gg |\alpha_b|$, so that
one can take $E_b[{\cal E}(t)]\approx w_b$ and
$|b[{\cal E}(t)]\rangle \approx
|b\rangle$, and assume that only the upper state $|c\rangle$ is significantly
perturbed by the rf field. Moreover, we are also considering systems
for which $|b\rangle$ and $|c^{(1)}\rangle$ have the same parity and
therefore are not directly coupled to each other by the control laser field.
In these conditions, the time-dependent Schr\"odinger
equation for the rf-dressed system becomes
\begin{equation}
\frac{{d}\;}{{d}t}\begin{pmatrix} c_{b}\\c_{c}\end{pmatrix}=
\begin{pmatrix}0&
-{i}\Omega_{\text{c}}/2\\
-{i}\Omega_{\text{c}}/2&
{i}[\Delta_{\text{c}}-2\Sigma_c\sin^{2}\omega_{\text{rf}}t]\end{pmatrix}
\begin{pmatrix}c_{b}\\c_{c}\end{pmatrix}\label{eq:HT}
\end{equation}
where
\begin{equation}
\Sigma_c=\alpha_{c} {\cal E}_{\rm rf}^2/4.
\label{eq:StarkSigma}
\end{equation}

Eq.\ (\ref{eq:HT})
is a system of linear differential
equations with periodic coefficients,
which is amenable to the Floquet
description studied in the following section.
The effect of relaxation, which is not taken into account in 
Eq.\ (\ref{eq:HT}), will be considered in Sec.\
\ref{subsec:3b}.

%%%%%%%%%%%%%%%%%%%%%%%%%%%%%%%%%%%%%%%%%%%%%%%%%%%%%%%%%%%%%%%%%%%%%%%%%%%%%%%%%%%%%%%%%%%%%%%%%%%%%%%%%%%%%%%%%%%%%%%%%%%%%%%%%%%%%%%%%%%%%%%%%%%%%%%%%%%%%%%%%%%%%%%%%%%%%%%%%%%%%%%%%%%%%%%%%%%%%%%%%%%%%%%%%%%%%%%%%%%%%%%%%%%%%%%%%%%%%%%%%%%%%%%%%%%%%%%%%%%%%%%%%%%%%%%%%%%%%%%%%%%%%%%%%%%%%%%%%%%%%%%%
\subsection{Floquet formalism}\label{subsec:3a}
\subsubsection{Floquet states}\label{subsubsec:3a1}
%%%%%%%%%%%%%%%%%%%%%%%%%%%%%%%%%%%%%%%%%%%%%%%%%%%%%%%%%%%%%%%%%%%%%%%%%%%%%%%%%%%%%%%%%%%%%%%%%%%%%%%%%%%%%%%%%%%%%%%%%%%%%%%%%%%%%%%%%%%%%%%%%%%%%%%%%%%%%%%%%%%%%%%%%%%%%%%%%%%%%%%%%%%%%%%%%%%%%%%%%%%%%%%%%%%%%%%%%%%%%%%%%%%%%%%%%%%%%%%%%%%%%%%%%%%%%%%%%%%%%%%%%%%%%%%%%%%%%%%%%%%%%%%%%%%%%%%%%%%%%%%%
By virtue of the Floquet
theorem~{ {\color{black}\cite{shirley,chu,barone,book}}}, 
any solution of Eq.\ (\ref{eq:HT})
can be written as a superposition of fundamental solutions
of the form 
%%%%%%%%%%%%%%%%%
%%%%%%%%%%%%%%%%%
\begin{subequations}
\label{eq:FSOL}
\begin{eqnarray}
{c_b(\epsilon_{};t)}&=&{e}^{-{i}\epsilon_{}t}\sum_{n=-\infty}^{\infty}
c_{b,n}
{e}^{-2{i}n\omega_{\text{rf}}t},
\label{eq:FSOL1} \\
{c_c(\epsilon_{};t)}&=&{e}^{-{i}\epsilon_{}t}\sum_{n=-\infty}^{\infty}
c_{c,n}
{e}^{-2{i}n\omega_{\text{rf}}t},
\label{eq:FSOL2}
\end{eqnarray}
\end{subequations}
%%%%%%%%%%%%%%%%%
%%%%%%%%%%%%%%%%%
where the quasienergy $\epsilon_{}$
and the coefficients $c_{b,n}$ and $c_{c,n}$ are constants.
(The fundamental angular frequency is $2\omega_{\rm rf}$, not
$\omega_{\rm rf}$, because the time-dependence entirely arises from the periodic
Stark shift of the upper level, which varies at twice the rf frequency.)
Replacing 
$c_b(t)$ and $c_c(t)$ by ${c_b(\epsilon_{};t)}$ and ${c_c(\epsilon_{};t)}$ turns
Eq.\ (\ref{eq:HT}) into a time-independent system of algebraic equations
for the coefficients $c_{b,n}$ and $c_{c,n}$,
namely
%%%%%%%%%%%%%%%%%
%%%%%%%%%%%%%%%%%
\begin{subequations}
\label{eq:CEQ}
\begin{flalign}
\frac{\Omega_{\text{c}}}{2}c_{c,n}-2n\omega_{\text{rf}}c_{b,n}& =  \epsilon_{}c_{b,n},\label{eq:CEQ1}\\
(\Sigma_c-\Delta_{\text{c}}-2n\omega_{\text{rf}})c_{c,n}+\frac{\Omega_{\text{c}}}{2}c_{b,n} & \nonumber\\
-\frac{\Sigma_c}{2}(c_{c,n-1}+c_{c,n+1}) & =\epsilon_{}c_{c,n}.\label{eq:CEQ2}
\end{flalign}
\end{subequations}
%%%%%%%%%%%%%%%%%
%%%%%%%%%%%%%%%%%
The quasienergies are those values of $\epsilon_{}$ for which this system has
a non-trivial solution.
In matrix form, Eqs.\ (\ref{eq:CEQ1}) and (\ref{eq:CEQ2}) read
%%%%%%%%%%%%%%%%%
%%%%%%%%%%%%%%%%%
%%\begin{equation}
%%\hat{H}_{2{\rm F}}\ket{\Psi}_{k}=\epsilon_{k}\ket{\Psi}_{k}.\label{eq:MF}
%%\end{equation}
\begin{equation}
\mathsf{H}_{2{\rm F}}\,\mathsf{c}=\epsilon_{}\mathsf{c},
\label{eq:MF}
\end{equation}
%%%%%%%%%%%%%%%%%
%%%%%%%%%%%%%%%%%
where $\mathsf{c}$ is the column vector
$(\ldots,c_{b,-1},c_{b,0},c_{b,+1},\ldots,c_{c,-1},c_{c,0},c_{c,+1},\ldots)^{T}$ and
%%where $\ket{\Psi}_{k}$ is the infinite 
%%column vector whose components are 
%%the harmonic components $b_{n}$ and $c_{n}$ i.e. 
%%$\ket{\Psi}_{k}=(\ldots,b_{-1},b_{0},b_{+1},\ldots,c_{-1},c_{0},c_{+1},\ldots)^{T}$ 
%%and $\hat{H}_{2{\rm F}}$ is known as Floquet Hamiltonian, 
%%which is an infinite Hermitian matrix. 
%%If we define the Floquet state, 
%%$\ket{\phi,n}$, 
%%where $\phi$ represents the atomic state and 
%%$n$ is the harmonic component of Fourier expansion, 
%%the state vector $\ket{\Psi}_{k}$ is
%%%%%%%%%%%%%%%%%
%%%%%%%%%%%%%%%%%
%%\begin{equation}
%%\ket{\Psi}_{k}=\sum_{m=-\infty}^{\infty}\sum_{\phi=b,c}\phi_{m}\ket{\phi,m}.\label{eq:PSI}
%%\end{equation}
%%%%%%%%%%%%%%%%%
%%%%%%%%%%%%%%%%%
%%Also the matrix representation of $\hat{H}_{2{\rm F}}$, in this basis set, is,
%%%%%%%%%%%%%%%%%
%%%%%%%%%%%%%%%%%
\begin{widetext}
\begin{equation}\label{eq:MHF1}
\mathsf{H}_{2{\rm F}}=\begin{pmatrix}\ddots & \vdots & \vdots & \vdots & \vdots & \vdots & \vdots & \vdots & \\
                              \cdots & 2\omega_{\text{rf}} & 0 & 0 & \cdots &\Omega_{\text{c}}/2 & 0 & 0 & \cdots \\
                              \cdots & 0 & 0 & 0 & \cdots & 0 & \Omega_{\text{c}}/2 & 0 & \cdots \\ 
                              \cdots & 0 & 0 & -2\omega_{\text{rf}} & \cdots & 0 & 0 & \Omega_{\text{c}}/2 & \cdots \\
                              \cdots & \vdots & \vdots& \vdots & \cdots & \vdots & \vdots & \vdots & \cdots\\
                              \cdots & \Omega_{\text{c}}/2 & 0 & 0 & \cdots &\Sigma_c-\Delta_{\text{c}}+2\omega_{\text{rf}} & -\Sigma_c/2 & 0 & \cdots \\
                              \cdots & 0 & \Omega_{\text{c}}/2 & 0 &\cdots & -\Sigma_c/2 & \Sigma_c-\Delta_{\text{c}} & -\Sigma_c/2 & \cdots\\
                              \cdots & 0 & 0 & \Omega_{\text{c}}/2 &\cdots & 0 & -\Sigma_c/2 & \Sigma_c-\Delta_{\text{c}}-2\omega_{\text{rf}} & \cdots \\
                                     & \vdots & \vdots & \vdots & \vdots &\vdots & \vdots & \vdots & \ddots\end{pmatrix}.
\end{equation}
\end{widetext}
%%%%%%%%%%%%%%%%%
%%%%%%%%%%%%%%%%%
The possible values of the quasienergy are
the eigenvalues of $\mathsf{H}_{2{\rm F}}$, which we denote by $\epsilon_{k}$.
Each one corresponds to
a particular dressed state of the system, i.e., to
a solution of the Schr\"odinger
equation of
the form
\begin{equation}
|\Psi_k(t)\rangle = 
{e}^{-{i}\epsilon_{k}t}\sum_{n=-\infty}^{\infty}
{e}^{-2{i}n\omega_{\text{rf}}t} |\psi_{k,n}\rangle,
\end{equation}
where the state vectors $|\psi_{k,n}\rangle$ are time-independent. It follows
from the above that
$|\Psi_k(t)\rangle = c_b(\epsilon_k;t)|b\rangle +  c_c(\epsilon_k;t)|c\rangle$.
We thus have two equivalent descriptions of the dressed states of the system,
namely one in terms of the time-dependent coefficients
$c_b(\epsilon_k;t)$ and
$c_c(\epsilon_k;t)$ and one in terms of 
the time-independent vector
$|\Psi_k\rrangle\equiv(|\psi_{k,0}\rangle,|\psi_{k,\pm 1}\rangle,|\psi_{k,\pm 2}\rangle,
\ldots)^{{T}}$ formed by the harmonic components of $|\Psi_k(t)\rangle$.
(Here and in the following, we use the notation $| \ldots \rrangle$ to
represent a column vector of state vectors.)
The vector $\mathsf{c}_k$, solution of Eq.\ (\ref{eq:MF}) with
$\epsilon=\epsilon_k$, is the representation of $|\Psi_k\rrangle$
in the basis of the bare Floquet states 
\begin{equation}
|b,n\rrangle\equiv
(|b\rangle\, \delta_{in},i=0,\mp 1, \mp 2, \ldots)^{{T}}
\label{eq:repb}
\end{equation}
and
\begin{equation}
|c,n\rrangle\equiv
(|c\rangle\, \delta_{in},i=0,\mp 1, \mp 2, \ldots)^{{T}}.
\label{eq:repc}
\end{equation}
Thus \cite{notesigns}
\begin{equation}
|\Psi_k\rrangle=\sum_{n=-\infty}^\infty (c_{b,-n}|b,n\rrangle
+c_{c,n}|c,-n\rrangle).
\label{eq:Psiexp}
\end{equation}

The Hamiltonian matrix $\mathsf{H}_{2{\rm F}}$ has a four-block structure.
Its upper and lower diagonal blocks,
$\mathsf{H}_{b}$ and $\mathsf{H}_{c}$ respectively, 
 are square matrices
representing the Hamiltonians
of the Floquet manifolds spawned by the bare states $|b\rangle$ and $|c\rangle$.
The non-zero elements of the off-diagonal blocks arise from
the coupling of the states $|b\rangle$ and $|c\rangle$ by the control field, while
the off-diagonal elements $-\Sigma_c/2$ of $\mathsf{H}_{c}$ arise
from the coupling of the state $|c\rangle$ with itself via the absorption or stimulated
emission of two rf photons. The rf field thus mixes the different vectors
$|c,n\rrangle$ with each other. By contrast,
the matrix $\mathsf{H}_{b}$ is diagonal since we neglect
the dressing of the state $|b\rangle$ by the rf field. 
%(When acting together
%with the control laser field, however, the rf field also
%mixes the states $|b,n\rangle$ with each other
%through diagrams involving the absorption of a laser photon, the exchange of
%several rf photons, and the stimulated emission of a laser photon.)

It is useful to diagonalize the
matrix $\mathsf{H}_{c}$, which can be done
by transforming the basis from
$\{|b,n\rrangle,|c,n\rrangle,\, n=0,\pm 1, \pm 2,\ldots\}$ to
$\{|b',n\rrangle,|c',n\rrangle,\, n=0,\pm 1, \pm 2,\ldots\}$, with
$\kket{b',n}\equiv\kket{b,n}$ and
%%%%%%%%%%%%%%%%%
%%%%%%%%%%%%%%%%%
\begin{equation}
\kket{c',n}=\sum_{m=-\infty}^{\infty}J_{n-m}\left(
\frac{\Sigma_c}{2\omega_{\text{rf}}}
\right)\kket{c,m},\label{eq:NEWEIGEN2}
\end{equation}
%%%%%%%%%%%%%%%%%
%%%%%%%%%%%%%%%%%
where $J_{p}(x)$ is the $p$-th order 
Bessel function of the first kind \cite{chu}.
Under this change of basis, the matrix representing the Hamiltonian becomes 
%%%%%%%%%%%%%%%%%
%%%%%%%%%%%%%%%%%
\begin{widetext}
\begin{equation}\label{eq:MHF2}
\mathsf{H}_{2{\rm F}}'=\begin{pmatrix}\ddots & \vdots & \vdots & \vdots & \vdots & \vdots & \vdots & \vdots & \\
                              \cdots & 2\omega_{\text{rf}} & 0 & 0 & \cdots &\Omega_{\text{c}} J_0/2 & \Omega_{\text{c}} J_{+1}/2 & \Omega_{\text{c}} J_{+2}/2 & \cdots \\
                              \cdots & 0 & 0 & 0 & \cdots & \Omega_{\text{c}} J_{-1}/2 & \Omega_{\text{c}} J_{0}/2 & \Omega_{\text{c}} J_{+1}/2 & \cdots \\ 
                              \cdots & 0 & 0 & -2\omega_{\text{rf}} & \cdots & \Omega_{\text{c}} J_{-2}/2 & \Omega_{\text{c}} J_{-1}/2 & \Omega_{\text{c}} J_{0}/2 & \cdots \\
                              \cdots & \vdots & \vdots& \vdots & \cdots & \vdots & \vdots & \vdots & \cdots\\
                              \cdots & \Omega_{\text{c}} J_{0}/2 & \Omega_{\text{c}} J_{-1}/2 & \Omega_{\text{c}} J_{-2}/2 & \cdots &\Sigma_c-\Delta_{\text{c}}+2\omega_{\text{rf}} & 0 & 0 & \cdots \\
                              \cdots & \Omega_{\text{c}} J_{+1}/2 & \Omega_{\text{c}} J_{0}/2 & \Omega_{\text{c}} J_{-1}/2 &\cdots & 0 & \Sigma_c-\Delta_{\text{c}} & 0 & \cdots\\
                              \cdots & \Omega_{\text{c}} J_{+2}/2 & \Omega_{\text{c}} J_{+1}/2 & \Omega_{\text{c}} J_{0}/2 &\cdots & 0 & 0 & \Sigma_c-\Delta_{\text{c}}-2\omega_{\text{rf}} & \cdots \\
                                     & \vdots & \vdots & \vdots & \vdots &\vdots & \vdots & \vdots & \ddots\end{pmatrix},
\end{equation}
\end{widetext}
%%%%%%%%%%%%%%%%%
%%%%%%%%%%%%%%%%%
where $J_{p}$, $p=0,\pm 1,\ldots$,
denotes $J_{p}(\Sigma_c/2\omega_{\text{rf}})$. 
Neither the quasienergies nor
the diagonal elements of the Hamiltonian matrix
are affected by this change of basis.
We see that in the absence of the control field, i.e., for $\Omega_{\text{c}}=0$,
the vectors  $|b',n\rrangle$ and $|c',m\rrangle$ describe the Floquet sidebands
of the states $|b\rangle$ and $|c\rangle$, respectively. When $\Omega_{\text{c}}\not=0$,
the sidebands of $b$ are coupled to those of $c$ 
by the off-diagonal blocks of $\mathsf{H}_{2{\rm F}}'$, the strength
of the coupling between the $n$-th side band of $b$ and the $m$-th side band
of $c$ being proportional to $|J_{m-n}(\Sigma_c/2\omega_{\text{rf}})|$
\cite{townes}.

\subsubsection{Quasienergy spectrum}\label{subsubsec:3a2}

\begin{figure}[!t]
\begin{center}
\includegraphics*[width=8.5cm]{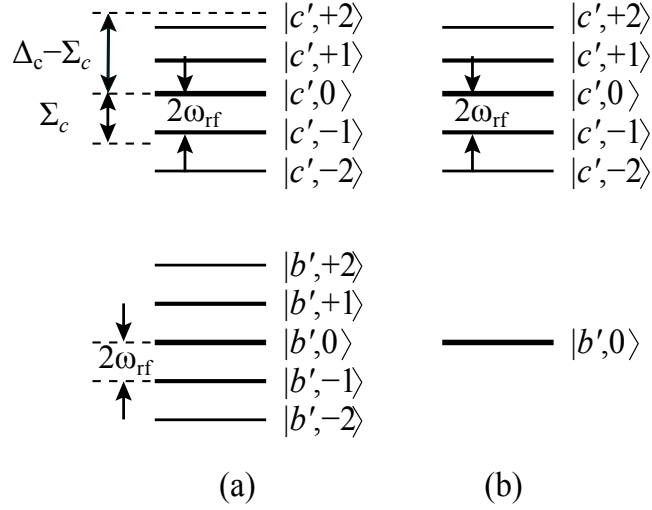}
\caption{\label{fig:Fig2}
(a) The manifold structure of the dressed 2-level system. The energy separation
between adjacent sidebands is $2\omega_{\rm rf}$. The energy of 
state $|c\rangle$ is Stark shifted
by the rf field by an amount $\Sigma_c$. The effective detuning from resonance
of the laser field coupling the two states is thus
$\Delta_{\rm c}-\Sigma_c$.
(b) The manifold structure in the $N$-level approximation: the sidebands of
state $|b\rangle$ are neglected and the interaction between the two manifolds
reduces to the interaction of the single state $|b\rangle$ with the $N$ states forming
the $c$ manifold.
}
\end{center}
\end{figure}
$\mathsf{H}_{2{\rm F}}$ and
$\mathsf{H}_{2{\rm F}}'$ reduce to the same diagonal matrix in the absence of control field, i.e.,
for $\Omega_{\rm c}=0$.
It follows from Eq.\ (\ref{eq:MHF2}) that their spectrum is a
double comb of quasienergies:
When $\Omega_{\rm c}=0$, any quasienergy $\epsilon_k$ is equal either to
$\epsilon_{b,n}^{(0)}$ or to $\epsilon_{c,n}^{(0)}$ for some value of
$n$, where
\begin{subequations}
\begin{eqnarray}
\epsilon_{b,n}^{(0)} &=& 2n \omega_{\text{rf}}, \\
\epsilon_{c,n}^{(0)} &=& \Sigma_c-\Delta_{\text{c}} + 2n \omega_{\text{rf}}.
\end{eqnarray}
\end{subequations}
The corresponding dressed states, $|\Psi_k\rrangle$, are
either $|b',n\rrangle$ or $|c',n\rrangle$, respectively.
Each of the
quasienergies is thus associated with an energy sideband of either state $|b\rangle$
or state $|c\rangle$. The sidebands of $|b\rangle$ are located at the energies
$E_{b,n}^{(0)}= w_b+\epsilon_{b,n}^{(0)}=w_b+2n\omega_{\text{rf}}$,
$n=0,\pm 1,\ldots$, 
and those of $|c\rangle$ at the energies
$E_{c,n}^{(0)}= w_b+\omega_{\text{c}}+\epsilon_{c,n}^{(0)}=
w_c+\Sigma_c+2n\omega_{\text{rf}}$,
$n=0,\pm 1,\ldots$ --- see Fig.\ \ref{fig:Fig2}(a).
As should be expected when $\Omega_{\rm c}=0$,
the sideband energies $E_{b,n}^{(0)}$ and
$E_{c,n}^{(0)}$ 
do not depend on the frequency
of the control laser. However, the quasienergies of the
$\epsilon_{c,n}^{(0)}$-manifold of quasienergies change
with the detuning $\Delta_{\text{c}}$,
with the consequence that this manifold
crosses the ($\omega_{\text{c}}$-independent)
$\epsilon_{b,n}^{(0)}$-manifold at 
$\Delta_{\text{c}}=\Sigma_c+2m\omega_{\text{rf}}$,
$m=0,\pm 1, \pm 2,\ldots$

\begin{figure}[!t]
\begin{center}
\includegraphics*[width=8.5cm]{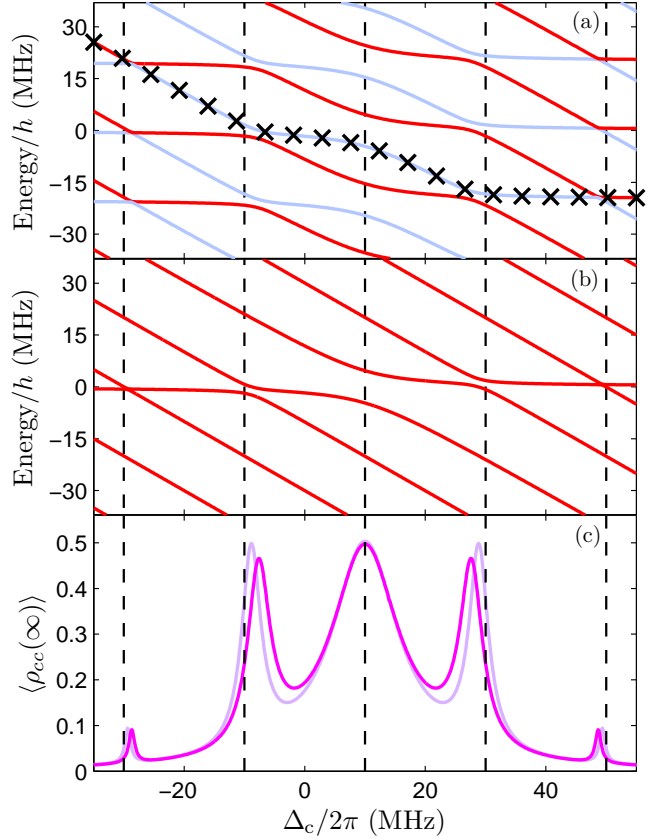}
\caption{\label{fig:Fig3}
(color online)
The quasienergy spectrum, (a) and (b), and the average population in the
upper state at large times ($t \gg 1/\Gamma_c$, as defined
in Section \ref{subsec:3b}), (c),
in an rf-dressed 2-level system.
$\Omega_{\text{c}}/2\pi=10$ MHz, $\Sigma_c/2\pi=10$ MHz,
$\omega_{\text{rf}}/2\pi=10$ MHz,
$\Gamma_{{c}}/2\pi=1$ MHz and $\gamma_{\rm c}=0$.
(a) and dark pink curve in (c): exact results.
(b) and light purple curve in (c):
results obtained in the $N$-level approximation.
Maxima in the population transferred to the upper state occur 
at the detunings at which the quasienergy curves cross or anticross each
other. Differences in the positions and amplitudes of these maxima between
the exact results and the $N$-level approximation are visible in (c).
The crosses represent one of the quasienergy curves predicted by
Eqs.\ (\ref{eq:QEPN1}) and (\ref{eq:QEPN2}).
}
\end{center}
\end{figure}
The quasienergy spectrum keeps that double comb structure for 
$\Omega_{\rm c}\not=0$: any quasienergy $\epsilon_k$ is equal either to
$\epsilon_{b,n}$ or to $\epsilon_{c,n}$ for some value of $n$, with
\begin{subequations}
\begin{eqnarray}
\epsilon_{b,n}&=&\epsilon_{b,0}+2n\omega_{\text{rf}}, \\
\epsilon_{c,n}&=&\epsilon_{c,0}+2n\omega_{\text{rf}}, 
\end{eqnarray}
\end{subequations}
$\epsilon_{b,0}$ and $\epsilon_{c,0}$ being such that
$\epsilon_{b,0}\rightarrow\epsilon_{b,0}^{(0)}$ and 
$\epsilon_{c,0}\rightarrow\epsilon_{c,0}^{(0)}$ 
in the limit $\Omega_{\text{c}}\rightarrow 0$.
Thus $\epsilon_{b,n}\approx\epsilon_{b,n}^{(0)}$
and $\epsilon_{c,n}\approx\epsilon_{c,n}^{(0)}$ if the control field is 
sufficiently weak.
However, 
when $\Omega_{\text{c}}\not=0$,
the $\epsilon_{b,n}$-manifold and the
$\epsilon_{c,n}$-manifold are no longer degenerate at 
$\Delta_{\text{c}}=\Sigma_c+2m\omega_{\text{rf}}$,
$m=0,\pm 1, \pm 2,\ldots$\ 
Instead, as illustrated by
{{\color{black}Fig.\ \ref{fig:Fig3}(a)}},
these two manifolds anticross each
other in sequences of avoided crossings.

The gaps and positions of these avoided crossing are determined by
the off-diagonal blocks of the Hamiltonian matrix
$\mathsf{H}_{2{\rm F}}'$. If $\Sigma_c=0$, i.e., if the rf field is vanishingly weak,
$J_p(\Sigma_c/2\omega_{\text{rf}})=\delta_{p0}$, and these off-diagonal
blocks are the unit matrix multiplied by $\Omega_{\text{c}}/2$. 
In this case, the system is equivalent to an infinite set of non-interacting
two-level systems. One then has
\begin{subequations}
\begin{eqnarray}
\epsilon_{b,n}&=&-\frac{\Delta_{\text{c}}}{2}\pm\frac{1}{2}\sqrt{\Delta_{\text{c}}^2+\Omega_{\text{c}}^2}+2n\omega_{\text{rf}}, 
\label{eq:QEP1} \\
\epsilon_{c,n}&=&-\frac{\Delta_{\text{c}}}{2}\mp\frac{1}{2}\sqrt{\Delta_{\text{c}}^2+\Omega_{\text{c}}^2}+2n\omega_{\text{rf}},
\label{eq:QEP2} 
\end{eqnarray}
\end{subequations}
where the upper signs apply
for $\Delta_{\text{c}} > 0$ and the lower signs for
$\Delta_{\text{c}} < 0$. The states interchange character at
$\Delta_{\text{c}} = 0$. However, since in the absence of the rf field
Floquet harmonic components of 
different $n$ values are not coupled to each other,
the crossings between the $b$ and the $c$
manifolds at
$\Delta_{\text{c}} = 2m\omega_{\text{rf}}$ ($m\not=0$) are true crossings
rather than avoided crossings.

The off-diagonal elements of the off-diagonal blocks are normally non-zero
if the rf field is not vanishingly weak,
but they rapidly decrease away
from the diagonal when
$|\Sigma_c/2\omega_{\text{rf}}| \ll 1$ since 
$J_p(\Sigma_c/2\omega_{\text{rf}})$ then goes rapidly to zero when $|p|$ increases.
If one neglects these off-diagonal elements,
Eqs.\ (\ref{eq:QEP1}) and (\ref{eq:QEP2})
then generalize to
$\epsilon_{b,n}\approx\epsilon_{b,n}^{(1)}$
and $\epsilon_{c,n}\approx\epsilon_{c,n}^{(1)}$, with
%%%%%%%%%%%%%%%%%
%%%%%%%%%%%%%%%%%
%%\begin{equation}
\begin{flalign}
\epsilon_{b,n}^{(1)}&=\frac{1}{2}
\left(\epsilon_{b,n}^{(0)}+\epsilon_{c,n}^{(0)}\right)
\pm\frac{1}{2}\sqrt{
\left(\epsilon_{b,n}^{(0)}-\epsilon_{c,n}^{(0)}\right)^2
+\Omega_{\text{c}}^2 J_{0}^2} \nonumber \\
&=\frac{1}{2}
\left(\Sigma_c-\Delta_{\text{c}}\right)
\pm\frac{1}{2}\sqrt{\left(\Sigma_c-\Delta_{\text{c}}\right)^2
+\Omega_{\text{c}}^2 J_{0}^2} + 2 n\omega_{\text{rf}}
\label{eq:QENERN1} 
\end{flalign}
%%\end{equation}
and
%%\begin{equation}
\begin{flalign}
\epsilon_{c,n}^{(1)}&=\frac{1}{2}
\left(\epsilon_{b,n}^{(0)}+\epsilon_{c,n}^{(0)}\right)
\mp\frac{1}{2}\sqrt{
\left(\epsilon_{b,n}^{(0)}-\epsilon_{c,n}^{(0)}\right)^2
+\Omega_{\text{c}}^2 J_{0}^2} \nonumber \\
&=\frac{1}{2}
\left(\Sigma_c-\Delta_{\text{c}}\right)
\mp\frac{1}{2}\sqrt{\left(\Sigma_c-\Delta_{\text{c}}\right)^2
+\Omega_{\text{c}}^2 J_{0}^2} + 2 n\omega_{\text{rf}}.
\label{eq:QENERN2}
\end{flalign}
%%\end{equation}
The upper signs apply for $\Delta_{\text{c}}> \Sigma_c$ and the lower
signs for $\Delta_{\text{c}}< \Sigma_c$. The control laser
field thus lifts the degeneracy between the states $|b',n\rangle$ and
$|c',n\rangle$ at $\Delta_{\text{c}}= \Sigma_c$ and creates an energy gap
approximately equal to $\Omega_{\text{c}}J_0(\Sigma_c/2\omega_{\text{rf}})$
between $\epsilon_{b,n}$ and $\epsilon_{c,n}$.
In 
{{\color{black}Fig.\ \ref{fig:Fig3}(a)}}, this interaction results in
conspicuous avoided crossings at $\Delta_{\text{c}}/(2\pi)\approx 10$ MHz.

The two manifolds formed by the quasienergy curves $\epsilon_{b,n}^{(1)}$
cross each other at the
detunings at which
$\epsilon_{b,n}^{(1)}=\epsilon_{c,n\pm m }^{(1)}$, $m=1, 2,\ldots$,
i.e., at $\Delta_{\text{c}}=\Delta_{\text{c},\pm m}$ with
\begin{equation}\label{eq:SIDEBAND}
\Delta_{\text{c},\pm m}=
\Sigma_c\pm\sqrt{4m^2\omega_{\text{rf}}^2-\Omega_{\text{c}}^2J_0^2}.
\end{equation}
Therefore these true crossings do not occur exactly at
$\Delta_{\text{c}}=\Sigma_c\pm 2m\omega_{\text{rf}}$,
as would be the case if $\Omega_{\text{c}}=0$,
but instead at detunings shifted towards the avoided
crossings at $\Delta_{\text{c}}=\Sigma_c$. Expanding the square root function
to lowest order in $\Omega_{\text{c}}$, Eq.\ (\ref{eq:SIDEBAND}) gives
\begin{equation}
\label{eq:SIDEBAND2}
\Delta_{\text{c},\pm m}\approx \Sigma_c\pm 2m\omega_{\text{rf}}\mp
\Omega_{\text{c}}^2 J_{0}^2/(4m\omega_{\text{rf}}). 
\end{equation}
This result can also be obtained using perturbation theory.

The degeneracy of the $\epsilon_{b,n}^{(1)}$
and $\epsilon_{c,n\pm m}^{(1)}$ quasienergies at $\Delta_{\text{c},\pm m}$
is lifted by the off-diagonal terms of the off-diagonal blocks
of the Hamiltonian matrix $\mathsf{H}_{2{\rm F}}'$. For instance,
the avoided crossings visible at $\Delta_{\text{c}}/(2\pi)\approx -10$ MHz
and 30 MHz in 
{{\color{black}Fig.\ \ref{fig:Fig3}(a)}} originate from the terms
$\Omega_{\text{c}}J_{\pm 1}$ in these
off-diagonal blocks, which couple the $|b',n\rrangle$ states to the
$|c',\pm 1\rrangle$ states.
Treating each pair of intersecting quasienergies as
an isolated two-level system in the vicinity of their intersection and
neglecting the terms in $\Omega_{\text{c}}J_{\pm m}$ with $m > 1$ in
$\mathsf{H}_{2{\rm F}}'$ yields the improved approximation 
$\epsilon_{b,n}\approx\epsilon_{b,n}^{(2)}$
and $\epsilon_{c,n}\approx\epsilon_{c,n}^{(2)}$, with
%%%%%%%%%%%%%%%%%
%%%%%%%%%%%%%%%%%
\begin{equation}
\epsilon_{b,n}^{(2)}=\frac{1}{2}
\left(\epsilon_{b,n}^{(1)}+\epsilon_{c,n-1}^{(1)}\right)
\pm\frac{1}{2}\sqrt{
\left(\epsilon_{b,n}^{(1)}-\epsilon_{c,n-1}^{(1)}\right)^2
+\Omega_{\text{c}}^2 J_{1}^2} \label{eq:QEPN1} 
\end{equation}
and
\begin{equation}
\epsilon_{c,n}^{(2)}=\frac{1}{2}
\left(\epsilon_{b,n+1}^{(1)}+\epsilon_{c,n}^{(1)}\right)
\mp\frac{1}{2}\sqrt{
\left(\epsilon_{b,n+1}^{(1)}-\epsilon_{c,n}^{(1)}\right)^2
+\Omega_{\text{c}}^2 J_{1}^2}. \label{eq:QEPN2}
\end{equation}
As in the above, the signs should be chosen so that 
$\epsilon_{b,n}^{(2)}\rightarrow \epsilon_{b,n}^{(0)}$ and
$\epsilon_{c,n}^{(2)}\rightarrow \epsilon_{c,n}^{(0)}$  for 
$\Omega_{\text{c}}\rightarrow 0$.
One of the resulting quasienergy curves is represented
by crosses in
{{\color{black}Fig.\ \ref{fig:Fig3}(a)}}. As seen from the figure, it reproduces the exact quasienergies well.
The only significant differences occur when
$\Delta_{\text{c}}\approx
\Delta_{\text{c},\pm m}$ with $m \geq 2$, where the approximate quasienergy
manifolds
$\epsilon_{b,n}^{(2)}$ and $\epsilon_{c,n}^{(2)}$ cross rather than anticross
each other. The approximation can be improved by taking into
account more off-diagonal components of the off-diagonal blocks of $\mathsf{H}_{2{\rm F}}'$
and further iterating the process leading 
%%from Eqs.\ (\ref{eq:QENERN}) to
to
Eqs.\ (\ref{eq:QEPN1}) and (\ref{eq:QEPN2}). The effective Rabi frequency
for two quasienergy curves intersecting
at $\Delta_{\text{c}}\approx
\Delta_{\text{c},\pm m}$ is 
$\Omega_{\rm c}|J_{\pm m}(\Sigma_c/2\omega_{\text{rf}})|$. The gap of the corresponding avoided
crossing is therefore a decreasing function of $m$ when,
as we have been assuming throughout this discussion,
$|\Sigma_c/(2\omega_{\text{rf}})| \ll 1$.

Treating the crossings at $\Delta_{\text{c}}\approx
\Delta_{\text{c},\pm 1}$ separately from that at $\Delta_{\text{c}}=
\Sigma_c$ is justified if the former are sufficiently far apart
from the latter, namely if $|\Delta_{\text{c},\pm 1}-\Sigma_c| \gg
\max\,(|\Omega_{\text{c}}J_0|,|\Omega_{\text{c}}J_1|)$. Given that
$J_0(\Sigma_c/2\omega_{\text{rf}}) \gg J_1(\Sigma_c/2\omega_{\text{rf}})$
when $\Sigma_c/(2\omega_{\text{rf}}) \ll 1$, this condition can also be written
%%%%%COND1%%%%%%
\begin{equation}\label{eq:COND1}
\left\lvert\frac{{\Omega_{\text{c}} J_0}}{{2}\,\omega_{\text{rf}}}\right\rvert\ll1.
\end{equation}
%%%%%%%%%%%%%%%
The iterative method outlined above fails
when the inequality (\ref{eq:COND1}) is violated. The crossings between
the manifolds may then occur at different values of $\Delta_{\text{c}}$
than predicted by Eqs.\ (\ref{eq:SIDEBAND}) and (\ref{eq:SIDEBAND2}).

Finally, we note that for weak fields
the quasienergies obtained by the iterative method
are consistent with those predicted by perturbation theory, namely, for
$n=0$ \cite{notechu},
\begin{subequations}
\begin{eqnarray}
\epsilon_{b,0}&\approx&-\sum_{n=-\infty}^\infty
{(\Omega_{\rm c} J_n/2)^2 \over \Sigma_c-\Delta_{\rm c}-2n\omega_{\rm rf}},
\label{eq:EPERT1} \\
\epsilon_{c,0}&\approx&\Sigma_c-\Delta_{\rm c}+\sum_{n=-\infty}^\infty
{(\Omega_{\rm c} J_n/2)^2 \over \Sigma_c-\Delta_{\rm c}-2n\omega_{\rm rf}}.
\label{eq:EPERT2}
\end{eqnarray}
\end{subequations}

%%%%%%%%%%%%%%%%%%%%%%%%%%%%%%%%%%%%%%%%%%%%%%%%%%%%%%%%%%%%%%%%%%%%%%%%%%%%%%%%%%%%%%%%%%%%%%%%%%%%%%%%%%%%%%%%%%%%%%%%%%%%%%%%%%%%%%%%%%%%%%%%%%%%%%%%%%%%%%%%%%%%%%%%%%%%%%%%%%%%%%%%%%%%%%%%%%%%%%%%%%%%%%%%%%%%%%%%%%%%%%%%%%%%%%%%%%%%%%%%%%%%%%%%%%%%%%%%%%%%%%%%%%%%%%%%%%%%%%%%%%%%%%%%%%%%%%%%%%%%%%%%
\subsubsection{The $N$-level approximation}\label{subsubsec:3a3}
%%If we are in the regime where 
%%{{\color{black}Eq.\ (\ref{eq:COND1})}} is satisfied,
%%the $N$-level picture is a good
%%approximation for the rf-dressed system.
%%The diagram, shown in { {\color{black}Fig.\ \ref{fig:Fig1}(b)}}, 
%%describes the $N$-level picture. It is composed of 
%%a ground state, $\ket{b',0}$, 
%%the manifold of the excited state, $\ket{c',n}$ and
%%the Rabi coupling between $\ket{b',0}$ and 
%%$\ket{c',n}$, given by $\Omega_{\text{c}} J_{n}$ 
%%from $\mathsf{H'}_{2{\rm F}}$ ;
%%thus the matrix form of the Hamiltonian of 
%%$N$-level system, $\mathsf{H}_{2N}$, is
As seen above, the bare state $|b\rangle$ turns into a manifold of sideband
states
when the system is described in the Floquet formalism. The
different states of this manifold are not directly coupled to each other 
by the rf field under our assumption
that the dressing of state $|b\rangle$ by this field
is negligible; however, they interact with each other indirectly,
through their
coupling with the harmonic components of state $|c\rangle$ by the
control laser field. Neglecting the manifold structure of state $|b\rangle$
amounts to setting $c_{b,n} \equiv 0$ for $n\not= 0$
in Eq.\ (\ref{eq:FSOL1}) and to
reducing the Hamiltonian matrix $\mathsf{H'}_{2{\rm F}}$ to the matrix
%%%%%%%%%%%%%%%
%%%%%%%%%%%%%%%
\begin{widetext}
%\begingroup
%\squeezetable
%\begin{table}[h]
\begin{equation}\label{eq:HN}
{\tilde{\sf H}}_{2{\rm F}}^{\prime}=\begin{pmatrix} 0 &  \cdots & \Omega_{\text{c}} J_{-1}/2 & \Omega_{\text{c}} J_{0}/2 & \Omega_{\text{c}} J_{+1}/2 & \cdots \\ 
                        \vdots&  \ddots & \vdots & \vdots & \vdots & \cdots\\
              \Omega_{\text{c}} J_{-1}/2 &  \cdots &\Sigma_c-\Delta_{\text{c}}+2\omega_{\text{rf}} & 0 & 0 & \cdots \\
               \Omega_{\text{c}} J_{0}/2 & \cdots & 0 & \Sigma_c-\Delta_{\text{c}} & 0 & \cdots\\
              \Omega_{\text{c}} J_{+1}/2 & \cdots & 0 & 0 & \Sigma_c-\Delta_{\text{c}}-2\omega_{\text{rf}} & \cdots \\
                       \vdots & \vdots &\vdots & \vdots & \vdots & \ddots\end{pmatrix}.
\end{equation}  
%\end{table}
%\endgroup
\end{widetext}
%%%%%%%%%%%%%%%
%%%%%%%%%%%%%%%
We refer to this approximation as the $N$-level approximation, in view
of the fact that
the matrix ${\tilde{\sf H}}_{2{\rm F}}^{\prime}$ is effectively the Hamiltonian
of a system consisting of a single state
$|b',0\rangle \equiv |b\rangle$
interacting with a manifold of $N$ independent states $|c',n\rrangle$,
$N\rightarrow\infty$ --- see Fig.\ \ref{fig:Fig2}(b).
As mentioned in the Introduction, this approximation
has often been made in studies of the interaction of low-lying states with
rf-dressed Rydberg states.

The corresponding quasienergies are the eigenvalues ${\tilde\epsilon}_k$ 
of the matrix ${\tilde {\sf H}}_{2{\rm F}}'$. An example of the resulting
spectrum is shown in Fig.\ \ref{fig:Fig3}(b).
Proceeding as in Sec.\ \ref{subsubsec:3a2}, one finds that for
$\Omega_{\text{c}}=0$ the corresponding quasienergies
are equal either to ${\tilde \epsilon}_b^{\,(0)}\equiv 0$ or
to ${\tilde \epsilon}_{c,n}^{\,(0)}\equiv 
\epsilon_{c,n}^{(0)}$ for some value of $n$, whereas for
$\Omega_{\text{c}}\not=0$ and $|\Sigma_c/(2\omega_{\text{rf}})| \ll 1$,
they are approximately equal either to ${\tilde \epsilon}_b^{\,(1)}\equiv
\epsilon_{b,0}^{(1)}$, to ${\tilde \epsilon}_{c,0}^{\,(1)}\equiv
\epsilon_{c,0}^{(1)}$, or to ${\tilde \epsilon}_{c,n}^{\,(1)}\equiv
\epsilon_{c,n}^{(0)} (n\not= 0)$. In the latter case,
the $c$-manifold intersects the
$b$-quasienergy curve at values of $\Delta_{\text{c}}$
for which ${\tilde \epsilon}_b^{\,(1)}
={\tilde \epsilon}_{c,n}^{\,(1)}$ for some value of $n$. However,
these intersections normally occur at different detunings than found in
Sec.\ \ref{subsubsec:3a2}, since in general
${\tilde \epsilon}_{c,n}^{\,(1)} \not= \epsilon_{c,n}^{(1)}$
when $n\not= 0$.
Instead, the crossings occur at $\Delta_{\text{c}}=
{\tilde \Delta}_{\text{c},\pm m}$
where 
\begin{equation}
\label{eq:SIDEBAND3}
{\tilde \Delta}_{\text{c},\pm m} = \Sigma_c \pm 2m\omega_{\text{rf}}
\mp \Omega_{\text{c}}^2 J_{0}^2/(8m\omega_{\text{rf}}).
\end{equation} 
Comparing this result with Eq.\ (\ref{eq:SIDEBAND2}), we see that within the
$N$-level approximation
the shift of the crossings from their zero-$\Omega_{\text{c}}$
positions is about half the correct value \cite{footnote2}.

%%%%%%%%%%%%%%%
%%%%%%%%%%%%%%%
%%\begin{equation}\label{eq:SHIFT}
%%\delta_{n}\approx\sum_{m\neq -n}\frac{(\Omega_{\text{c}} J_{m}/2)^2}{2\omega_{\text{rf}}(m+n)}+{\cal O}\left(\frac{\delta_{n}}{2\omega_{\text{rf}}(m+n)}\right),
%%\end{equation}
%%%%%%%%%%%%%%%
%%%%%%%%%%%%%%%
%%%%%%%%%%%%%%%
%%%%%%%%%%%%%%%
%%\begin{equation}\label{eq:FREQSHIFT}
%%[\mathsf{H}_{2N}]_{nn}\approx\Sigma_c-\Delta_{\text{c}}\pm2m\omega_{\text{rf}}\mp{\delta_{n}}.
%%\end{equation}
%%%%%%%%%%%%%%%
%%%%%%%%%%%%%%%

\subsection{Dressed state dynamics and relaxation}\label{subsec:3b}
%%\subsubsection{Full dynamics}\label{subsubsec:3b1}
Being entirely based on Eq.\ (\ref{eq:HT}),
the analysis developed in Sec.\ \ref{subsec:3a}
neglects both the frequency width of the control
laser and the energy width of the states $|b\rangle$ and $|c\rangle$.
The corresponding
time evolution of the system is purely oscillatory, with
population transferred to-and-fro between $|b\rangle$ and $|c\rangle$
without any damping. The evolution would 
be a pure Rabi flopping of characteristic angular frequency
$(\Delta_{\rm c}^2+\Omega_{\rm c}^2)^{1/2}$ in the absence of the rf field.
The oscillation is multimode in the presence of the latter, with 
Fourier components 
separated in angular frequency by integer multiples of $2\omega_{\rm rf}$.
{{\color{black}Eqs.\ (\ref{eq:QENERN1})}} and (\ref{eq:QENERN2}) approximately give
the corresponding characteristic frequencies as 
$[(\Sigma_c-\Delta_{\text{c}})^2+\Omega_{\text{c}}^2 J_{0}^2]^{1/2}/2 +
2p\omega_{\rm rf}$, $p=0,\pm 1,\pm 2,\ldots$

Relaxation modifies this picture.
Taking into account the natural
width of state $|c\rangle$, $\Gamma_c$, and the linewidth of the control laser,
$\gamma_{\rm c}$, and assuming that $|c\rangle$ can
de-excite only to $|b\rangle$,
the time evolution of the system is described by the
following optical Bloch equations:
\begin{subequations}
\begin{flalign}
\dot{\rho}_{bb}=&\; \Gamma_{{c}}\rho_{cc}+\frac{{i}\Omega_{\text{c}}}{2}(
\rho_{bc}-\rho_{cb}),\label{eq:HRHO1}\\
\dot{\rho}_{cc}=&-\Gamma_{{c}}\rho_{cc}+\frac{{i}\Omega_{\text{c}}}{2}(
\rho_{cb}-\rho_{bc}),\label{eq:HRHO2}\\
\dot{\rho}_{bc}=&-\left\lbrace\Gamma_c/2+\gamma_{\rm c}+{
 i}\left[\Delta_{\text{c}}-2\Sigma_c\sin^2\omega_{\text{rf}}t\right]\right\rbrace\
\rho_{bc}\nonumber\\
&+\frac{{i}\Omega_{\text{c}}}{2}(\rho_{bb}-\rho_{cc}).\label{eq:HRHO3}
\end{flalign}
\end{subequations}
(Recall that at the moment
we study the system formed by
states $|b\rangle$ and $|c\rangle$ in isolation from state $|a\rangle$.
We therefore assume, for the time being,
that $|b\rangle$ does not de-excite to a lower state.
The finite lifetime of state $|b\rangle$ will be taken into account
in the analysis of the full 3-level ladder system, in Sec.\
\ref{sec:4}.)
Decoherence dampens the dynamical evolution
of the system until the oscillation is purely
driven by the rf field. In this steady state, which is reached
for $t \gg 1/\Gamma_c$, the elements of
the density matrix oscillate at integer multiples of
the fundamental angular frequency 
$2\omega_{\text{rf}}$ about a constant cycle-average. 

In the $N$-level approximation, we treat each side band of state $|c\rangle$ as an
independent bound state and describe the evolution of the system by
optical Bloch equations based on the Hamiltonian matrix
${\tilde {\sf H}}_{2{\rm F}}'$, namely
\begin{subequations}
\begin{flalign}
\dot{\rho}_{bb}=& \;\Gamma_{{c}}\sum_k \rho_{kk}+
\frac{{i}\Omega_{\text{c}}}{2}
\sum_k (J_k\rho_{bk}-J_k\rho_{kb}),\label{eq:HRHON1}\\
\dot{\rho}_{bk}=& -\left[\Gamma_c/2+\gamma_{\rm c}-{
 i}(\Sigma_c-\Delta_{\text{c}}-2k\omega_{\rm rf})\right]
\rho_{bk}\nonumber\\
&+\frac{{i}\Omega_{\text{c}}}{2}\left(\rho_{bb}-\sum_jJ_j\rho_{jk}\right),\label{eq:HRHON2} \\
\dot{\rho}_{jk}=& -[\Gamma_{{c}}-2{i}(j-k)\omega_{\rm rf}]\rho_{jk}
+\frac{{i}\Omega_{\text{c}}}{2}(
J_k\rho_{jb}-J_j\rho_{bk}),\label{eq:HRHON3}
\end{flalign}
\end{subequations}
with the indexes $j$ and $k$ running over all the
states forming the $c$-manifold. In this approach,
the population in state $|c\rangle$, $\rho_{cc}$, is defined as $1-\rho_{bb}$,
or equivalently, as $\rho_{cc}=\sum_k \rho_{kk}$. In contrast to
Eqs.\ (\ref{eq:HRHO1}--\ref{eq:HRHO3}),
which predict oscillatory populations and coherences,
Eqs.\ (\ref{eq:HRHON1}--\ref{eq:HRHON3}) lead to
a constant density matrix in the
long time limit.

We assume that
the atom is initially in state $|b\rangle$ and
calculate the population transferred
to state $|c\rangle$, $\rho_{cc}(t)$, and its temporal average in the steady state,
$\langle\rho_{cc}(\infty)\rangle$. We define the latter as
%%%%%%%%%%%%%%%%%
\begin{equation}\label{eq:AV}
\langle\rho_{cc}(\infty)\rangle=\frac{1}{T}\lim_{t\to\infty}{\int_{t}^{t+T}{\rho
_{cc}(t')\,{d}t'}},
\end{equation}
%%%%%%%%%%%%%%%%%
with $T=2\pi/(2\omega_{\text{rf}})$.
We thus set $\rho_{bb}$ to 1 and 
all the other elements of the density matrix to zero
at time $t=0$ and numerically solve
either
Eqs.\ (\ref{eq:HRHO1}--\ref{eq:HRHO3}) or
Eqs.\ (\ref{eq:HRHON1}--\ref{eq:HRHON3}) 
from $t=0$
to $t \gg 1/\Gamma_c$. 

How the population transferred to state $|c\rangle$ varies
with the detuning $\Delta_{\rm c}$ depends both on the value
of $\Gamma_{c}$ and on whether
the inequality (\ref{eq:COND1}) is fulfilled or not.
Avoided crossings between quasienergy curves can be treated in isolation
when this inequality is fulfilled, and adjacent resonances do
not overlap significantly due to the natural width of the upper level
when $\Gamma_{c}\ll 2 \omega_{\rm rf}$.
The variation of $\langle\rho_{cc}(\infty)\rangle$
in these conditions is examplified by 
Fig.\ \ref{fig:Fig3}(c). The results represented respectively
by a dark pink
curve and by a light purple curve were obtained 
by solving 
Eqs.\ (\ref{eq:HRHO1}--\ref{eq:HRHO3}) and
Eqs.\ (\ref{eq:HRHON1}--\ref{eq:HRHON3}).

As seen from the figure, the avoided crossings in the dressed state spectrum
are associated with
strong enhancements of population transfer.
The origin of these enhancements is easily understood given that
$\Gamma_c$ and $\gamma_{\rm c}$ are both
sufficiently small that the evolution of the system is 
dominated by its dressed states dynamics rather than by relaxation.
For the parameters considered, the dressed atom can be described as a simple
two-state system in the vicinity of each avoided crossing. Sufficiently far
from any crossing, the dressed states are close to either one of the
$|b',n\rrangle$ states or one of the $|c',n\rrangle$ states, with little
admixture between the two. At a crossing, however,
the interacting dressed states are approximately equal
superpositions of $|b\rangle$ and 
$|c\rangle$. Population transfer is therefore enhanced at each crossing,
typically
over the range of values of $\Delta_{\rm c}$ for which the relevant
$b$- and $c$-dressed states
 are significantly admixed with each other.
Treating the dressed atom as a pure two-state system in the vicinity
of each crossing
leads to a simple 
expression for the population transferred to the upper state, i.e., for
$\Delta_{\rm c} \approx \Sigma_c+ 2 n \omega_{\rm rf}$ \cite{notechu},
%%%%%%%%%%%%%%%
%%%%%%%%%%%%%%%
\begin{equation}\label{eq:MODEL}
\langle\rho_{cc}(\infty)\rangle\approx{\Omega_{\text{c}}^2J_{n}^{2}/2 \over
\Gamma_c(\Gamma_c/2+{\gamma_{\rm c}})
[1+(\Delta_{\rm c}-\Sigma_c-2n\omega_{\rm rf})^2/\Gamma_c^2]+\Omega_{\text{c}}^2J_{n}^2}.
\end{equation}
%%%%%%%%%%%%%%%
%%%%%%%%%%%%%%%
Thus $\langle\rho_{cc}(\infty)\rangle$ exhibits power-broadened sidebands
whose height and width
are modulated by the square of the Bessel function factors $J_n$. When $|n|$ is
larger than the argument of these Bessel functions, these enhancements
decrease in amplitude and their width tends to the natural width of the upper
level, $\Gamma_c$, for increasing sideband orders.
For the parameters of Fig.\ \ref{fig:Fig3}(c),
this analysis predicts that the 
$n=0$, $n=\pm 1$ and
$n=\pm 2$ enhancements have an heigth of 0.50, 0.46 and 0.08, respectively,
and a full width at half maximum of 13.3, 3.6 and 1.1 MHz. As can be seen from the figure (dark purple curve),
these values are in agreement with the numerical results, although
the $n=0$ sideband somewhat
overlaps the others. (The
inequality (\ref{eq:COND1}) is fulfilled but marginally here, since
$|\Omega_{\rm c}J_0|/(2\omega_{\rm rf})\approx 0.5$.)

The picture is the same in the $N$-level approximation for the parameters
considered.
The discrepancy in the position of the enhancements mirrors the differences
in the position of the crossings in the quasienergy maps.
It can be largely eliminated by a simple $k$-dependent shift in the
$\Sigma_c-\Delta_{\rm c}-2k\omega_{\rm rf}$ factor multiplying $\rho_{bk}$ in
Eq.\ (\ref{eq:HRHON2}) \cite{footnote2}.

The comparison shows that for the parameters considered in Fig.\ \ref{fig:Fig3},
the upper state effectively turns into a manifold of states well
described within the $N$-level approximation.
As mentioned in {{\color{black}Sec.\ \ref{subsubsec:3a2}}}, increasing 
$\Omega_{\text{c}}$ eventually results in a break down of this approximation
as
the states forming the Floquet manifold will start to
interact with each other. For sufficiently strong control fields,
the resonance sidebands 
overlap each other and merge with the zeroth order resonance.

%%%%%%%%%%%%%%%%%%%%%%%%%%%%%%%%%%%%%%%%%%%%%%%%
%%%%%%%%%%%%%%%%%%%%%%%%%%%%%%%%%%%%%%%%%%%%%%%%
%%%%%%%%%%%%%%%%%%%%%%%%%%%%%%%%%%%%%%%%%%%%%%%%
%%%%%%%%%%%%%%%%%%%%%%%%%%%%%%%%%%%%%%%%%%%%%%%%
\begin{figure}[!t]
\begin{center}
\includegraphics*[width=10.5cm]{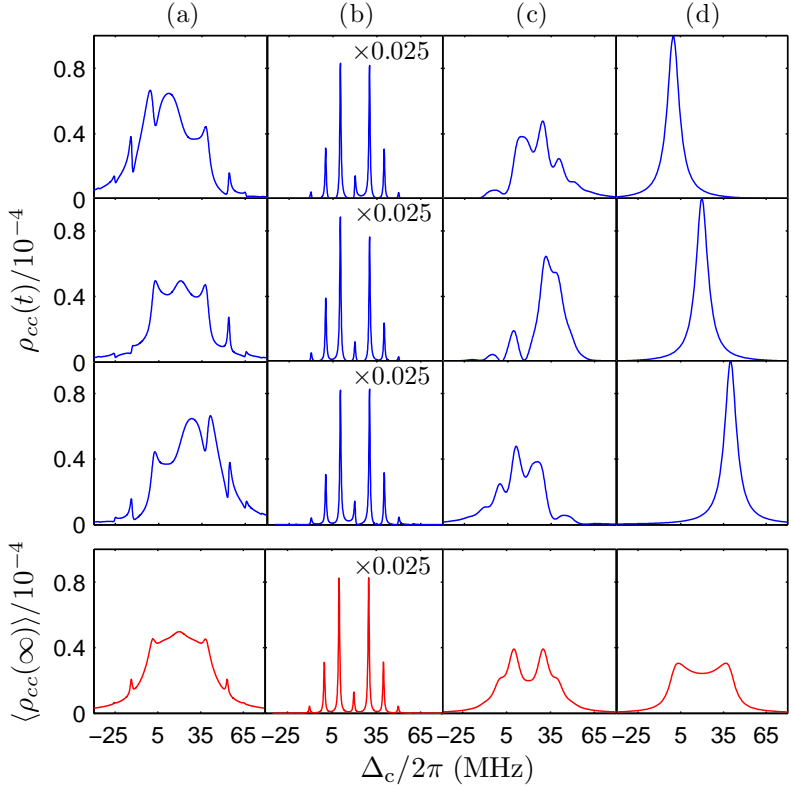}
\caption{\label{fig:Fig4}
(color online)
The population in the upper state at $\sin^2\omega_{\rm rf}t=0$ (first row),
$1/2$ (second row) and $1$ (third row), and its time average (bottom row),
in the steady state regime ($t \gg 1/\Gamma_c$).
$\Sigma_c/(2\pi)=20$ MHz and $\gamma_{\rm c}=0$ in all cases.
(a):  $\Omega_{\rm c}/2\pi= 20$ MHz, $\Gamma_c/2\pi=1$ MHz and
$\omega_{\rm rf}/2\pi=5$ MHz.
(b):  $\Omega_{\rm c}/2\pi= 0.1$ MHz, $\Gamma_c/2\pi=1$ MHz and
$\omega_{\rm rf}/2\pi=5$ MHz.
(c):  $\Omega_{\rm c}/2\pi= 0.1$ MHz, $\Gamma_c/2\pi=10$ MHz and
$\omega_{\rm rf}/2\pi=5$ MHz.
(d): $\Omega_{\rm c}/2\pi= 0.1$ MHz, $\Gamma_c/2\pi=10$ MHz and
$\omega_{\rm rf}/2\pi=0.01$ MHz.
$\Sigma_c/(2\pi)=20$ MHz and $\gamma_{\rm c}=0$ in all cases.
$\rho_{cc}$ is multiplied by a factor of 0.025 in column (b).
}
\end{center}
\end{figure}
%%%%%%%%%%%%%%%%%%%%%%%%%%%%%%%%%%%%%%%%%%%%%%%%
%%%%%%%%%%%%%%%%%%%%%%%%%%%%%%%%%%%%%%%%%%%%%%%%
%%%%%%%%%%%%%%%%%%%%%%%%%%%%%%%%%%%%%%%%%%%%%%%%
%%%%%%%%%%%%%%%%%%%%%%%%%%%%%%%%%%%%%%%%%%%%%%%%

How large $\lvert\Omega_{\text{c}} J_{0}\rvert$ is relative to
$2\omega_{\text{rf}}$ also impacts on the temporal evolution of the
system.
When $\lvert\Omega_{\text{c}} J_{0}\rvert\ll2\omega_{\text{rf}}$, 
the atom interacts with the laser field 
as if state $|c\rangle$ 
were a manifold of stationary states. 
In the opposite limit, 
the rf field changes slowly on the time scale on which the states $|b\rangle$ and $|c\rangle$
interact with each other. This interaction then occurs as if the rf field
was static at any instant. In this case, the Stark shift 
due to the rf field is expected to make the population in the state $|c\rangle$
oscillate periodically.

Columns (a) and (b) of
{ {\color{black}Fig.\ \ref{fig:Fig4}}} compare instantaneous and
cycle-average values of
$\rho_{cc}$ in the steady state regime for two different values of
$\Omega_{\text{c}}$. In column (a),
 $\lvert\Omega_{\text{c}} J_{0}\rvert/(2\omega_{\text{rf}})\approx 0.4$.
This value is insufficiently small for the $N$-state
picture to hold well, which leads to a clear oscillation in the population
in the state $|c\rangle$. In column (b),
 $\lvert\Omega_{\text{c}} J_{0}\rvert/(2\omega_{\text{rf}})\approx 0.002$,
which fulfills the condition (\ref{eq:COND2}): $\rho_{cc}$ is almost 
constant in time. The resonance sidebands are also much better marked 
than in column (a) because of the much smaller power broadening.

The population in the state $|c\rangle$ also depends on the
time scale of the relaxation mechanisms. Well defined enhancement sidebands
are not expected
unless spontaneous decay is slow compared to the oscillation of the
rf field, i.e., unless \cite{townes}
%%%%%%%%%%%%%%%
%%%%%%%%%%%%%%%
\begin{equation}\label{eq:COND2}
\Gamma_{{c}}\ll 2\omega_{\text{rf}}.
\end{equation}  
%%%%%%%%%%%%%%%
%%%%%%%%%%%%%%%
This inequality also expresses the requirement that adjacent 
sidebands, which are separated roughly by $2\omega_{\text{rf}}$, do not
overlap due to the
the natural linewidth of the resonance, $\Gamma_{{c}}$. The role 
of the lifetime of the state $|c\rangle$ is
illustrated by column (c) of 
{{\color{black}Fig.\ \ref{fig:Fig4}}}, which was calculated for the same
parameters as column (b) apart for $\Gamma_{{c}}$: while
$\Gamma_{{c}}\ll 2\omega_{\text{rf}}$ in column (b),
$\Gamma_{{c}}=2\omega_{\text{rf}}$ in column (c).
In the latter case the sidebands partly overlap each other and the population
in the state $|c\rangle$ oscillates markedly.

Finally, column (d) of
{ {\color{black}Fig.\ \ref{fig:Fig4}}} shows the time evolution of
$\rho_{cc}$ for the same parameters as in column (c) apart that
$2\omega_{\rm rf}$ is much smaller than both $\Gamma_{c}$ and
$\Delta_{\rm c}$. The $N$-level picture is now invalid and there is
no sign of sidebands. Instead, the profile of $\rho_{cc}$ is a single
peak of width $\Gamma_c$ following the instantaneous position of state $|c\rangle$
as this state is periodically Stark-shifted by the rf field \cite{townes}.
%%%%%%%%%%%%%%%%%%%%%%%%%%%%%%%%%%%%%%%%%%%%%%%%%%%%%%%%%%%%%%%%%%%%%%%%%%%%%%%%%%%%%%%%%%%%%%%%%%%%%%%%%%%%%%%%%%%%%%%%%%%%%%%%%%%%%%%%%%%%%%%%%%%%%%%%%%%%%%%%%%%%%%%%%%%%%%%%%%%%%%%%%%%%%%%%%%%%%%%%%%%%%%%%%%%%%%%%%%%%%%%%%%%%%%%%%%%%%%%%%%%%%%%%%%%%%%%%%%%%%%%%%%%%%%%%%%%%%%%%%%%%%%%%%%%%%%%%%%%%%%%%
%%%%%%%%%%%%%%%%%%%%%%%%%%%%%%%%%%%%%%%%%%%%%%%%%%%%%%%%%%%%%%%%%%%%%%%%%%%%%%%%%%%%%%%%%%%%%%%%%%%%%%%%%%%%%%%%%%%%%%%%%%%%%%%%%%%%%%%%%%%%%%%%%%%%%%%%%%%%%%%%%%%%%%%%%%%%%%%%%%%%%%%%%%%%%%%%%%%%%%%%%%%%%%%%%%%%%%%%%%%%%%%%%%%%%%%%%%%%%%%%%%%%%%%%%%%%%%%%%%%%%%%%%%%%%%%%%%%%%%%%%%%%%%%%%%%%%%%%%%%%%%%%

\section{Rf-dressed three-level systems}\label{sec:4}
\subsection{General formulation}\label{subsec:4a}

%%%%%%%%%%%%%%%%%%%%%%%%%%%%%%%%%%%%%%%%%%%%%%%%
\begin{figure}[!t]
\begin{center}
\includegraphics*[width=8.5cm]{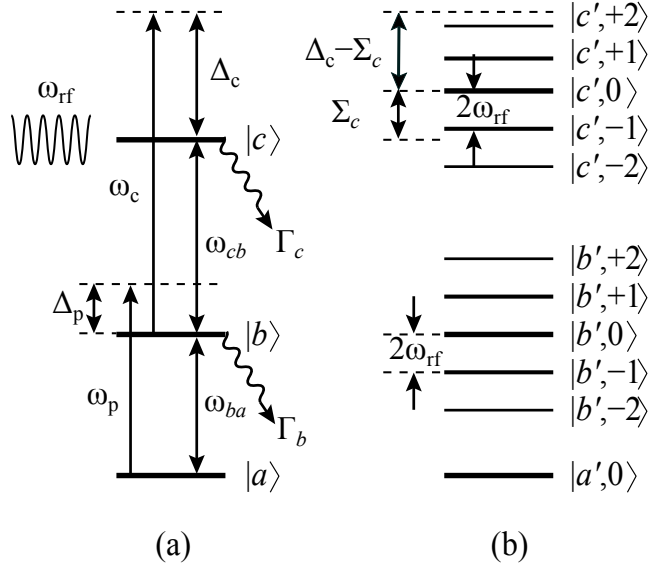}
\caption{\label{fig:Fig1}
(a) Schematic of the rf-dressed three-level 
Rydberg system.
The probe laser (angular frequency $\omega_{\text{p}}$)
couples the states $|{a}\rangle$ and $|{b}\rangle$,
while the control laser (angular frequency 
$\omega_{\text{c}}$) couples
the states $|{b}\rangle$ and $|{c}\rangle$. The energy of the
latter is modulated by the rf field (angular frequency
$\omega_{\rm rf}$).
The detunings of these two laser fields from
the respective transition frequencies are 
$\Delta_{\text{p}}$ and $\Delta_{\text{c}}$.
The state $|c\rangle$ decays to the state $|b\rangle$, and the state $|b\rangle$ to the state $|a\rangle$.
(b) The manifold structure of the states $|b\rangle$ and $|c\rangle$ induced
by the rf field.
The energy separation between 
adjacent sidebands is $2\omega_{\text{rf}}$. 
The rf field also shifts 
the energy of state $|{c}\rangle$ by an amount $\Sigma_c$;
the effective detuning of the control field is thus
$\Delta_{\text{c}}-\Sigma_c$.
}
\end{center}
\end{figure}
%%%%%%%%%%%%%%%%%%%%%%%%%%%%%%%%%%%%%%%%%%%%%%%%

We now take into account the
coupling of state $|b\rangle$ to state $|a\rangle$ by the probe laser field
and allow
state $|b\rangle$ to de-excite to state $|a\rangle$, which is assumed to be
stable (Fig.\ \ref{fig:Fig1}).
We proceed as in Sec.\ \ref{subsec:3a0} and, ignoring relaxation at first,
write the state vector of the 
atom as
\begin{equation}
|\Psi(t)\rangle = c_a(t)|a\rangle + c_b(t)|b\rangle + c_c(t)|c\rangle,
\end{equation}
with the amplitudes $c_a(t)$, $c_b(t)$, $c_c(t)$ satisfying the equation
\begin{equation}\label{eq:H3T}
{{d}\;\over{d}t}\begin{pmatrix} c_{a}\\c_{b}\\c_{c}\end{pmatrix}=\begin{pmatrix}0& -{i}\Omega_{\text{p}}/2&0\\ -{i}\Omega_{\text{p}}/2& {i}\Delta_{\text{p}}&-{i}\Omega_{\text{c}}/2\\0&-{i}\Omega_{\text{c}}/2&
{i}[\Delta_{\text{R}}-2\Sigma_c\sin^{2}\omega_{\text{rf}}t]\end{pmatrix}\begin{pmatrix}c_{a}\\c_{b}\\c_{c}\end{pmatrix}.
\end{equation}
Following the same analysis as in Sec.\ \ref{subsec:3a}, we formulate the
problem in terms of the Floquet Hamiltonian
%%%%%%%%%%%%%%%
%%%%%%%%%%%%%%%
\begin{equation}\label{eq:H3F}
\mathsf{H}_{3{\rm F}}=\begin{pmatrix}{\sf A} & \bm{\Omega_{\text{p}}} & {\sf 0} \\ \bm{\Omega_{\text{p}}}^{\dagger} & {\sf B} & \bm{\Omega_{\text{c}}} \\ {\sf 0} & \bm{\Omega_{\text{c}}}^{\dagger} & {\sf C} \end{pmatrix},
\end{equation}
%%%%%%%%%%%%%%%
%%%%%%%%%%%%%%%
where
\begin{subequations}
\begin{flalign}
{\sf A}= &\; \text{diag}(\ldots,2\omega_{\text{rf}},0,-2\omega_{\text{rf}},\ldots),\label{eq:CH3Fa}\\
{\bm\Omega_{\text{p}}}= &\; \text{diag}(\ldots,\frac{\Omega_{\text{p}}}{2},\frac{\Omega_{\text{p}}}{2},\frac{\Omega_{\text{p}}}{2},\ldots),\label{eq:CH3Fb}
\end{flalign}
\end{subequations}
and
\begin{equation}
\begin{pmatrix}
{\sf B} & \bm{\Omega_{\text{c}}} \\
\bm{\Omega_{\text{c}}}^{\dagger} & {\sf C} \end{pmatrix} =
\mathsf{H}_{2{\rm F}}-\Delta_{\rm p}
\begin{pmatrix}
{\sf 1} & {\sf 0} \\
{\sf 0} & {\sf 1} \end{pmatrix}.
\end{equation}
%%%%%%%%%%%%%%%
%%%%%%%%%%%%%%%
Generalizing the discussion of last section,
the corresponding Floquet states are described by 
time-independent vectors $|\Psi_k\rrangle$ which can be written in terms 
of the vectors $|a,n\rrangle \equiv
(|a\rangle\, \delta_{in},i=0,\mp 1, \mp 2, \ldots)^{{T}}$ and of the vectors
$|b,n\rrangle$ and $|c,n\rrangle$ defined by Eqs. (\ref{eq:repb}) and
(\ref{eq:repc}):
\begin{equation}
\label{eq:newpsi}
|\Psi_k\rrangle=\sum_{n=-\infty}^\infty
(c_{a,-n}|a,n\rrangle
+c_{b,-n}|b,n\rrangle
+c_{c,-n}|c,n\rrangle).
\end{equation}
As before, the ${\sf C}$-block of $\mathsf{H}_{3{\rm F}}$ can be diagonalized 
by a change of basis transforming the matrix $\mathsf{H}_{2{\rm F}}$ into
the matrix $\mathsf{H}_{2{\rm F}}'$ defined by Eq.\ (\ref{eq:MHF2}).
The change transforms the matrix $\mathsf{H}_{3{\rm F}}$ into
the matrix $\mathsf{H}_{3{\rm F}}'$ differing from  $\mathsf{H}_{3{\rm F}}$
by the replacement of $\mathsf{H}_{2{\rm F}}$ by
$\mathsf{H}_{2{\rm F}}'$. It does not affect the spectrum of quasienergy.
Making the $N$-level approximation further reduces $\mathsf{H}_{3{\rm F}}'$ to
\begin{widetext}
\begin{equation}\label{eq:H3N}
{\tilde{\sf H}}_{3{\rm F}}'=\begin{pmatrix} 0 & \Omega_{\text{p}}/2 & \cdots & 0 & 0 & 0 & \cdots \\
                            \Omega_{\text{p}}/2 & -\Delta_{\text{p}} & \cdots &
\Omega_{\text{c}} J_{-1}/2 & \Omega_{\text{c}} J_{0}/2 & \Omega_{\text{c}} J_{+1
}/2 & \cdots \\
                        0& \vdots&  \ddots & \vdots & \vdots & \vdots & \cdots\\
              0&\Omega_{\text{c}} J_{-1}/2 &  \cdots &\Sigma_c-\Delta_{\text{R}}+2
\omega_{\text{rf}} & 0 & 0 & \cdots \\
               0&\Omega_{\text{c}} J_{0}/2 & \cdots & 0 & \Sigma_c-\Delta_{\text{R
}} & 0 & \cdots\\
              0&\Omega_{\text{c}} J_{+1}/2 & \cdots & 0 & 0 & \Sigma_c-\Delta_{\text{R}}-2\omega_{\text{rf}} & \cdots \\
                \vdots&       \vdots & \vdots &\vdots & \vdots & \vdots & \ddots
\end{pmatrix}.
\end{equation}
\end{widetext}

The optical Bloch equations for the 3-state system with rf-modulation can
be written as
%%%%%%%%%%%%%%%%%
%%%%%%%%%%%%%%%%%
\begin{subequations}\label{eq:EMOTION}
\begin{flalign}
\dot{\rho}_{aa}=&\Gamma_{b}\rho_{bb}+\frac{{i}\Omega_{\text{p}}}{2}(\rho_{ab}-\rho_{ba}),\label{eq:EMOTION1}\\
\dot{\rho}_{bb}=&-\Gamma_{b}\rho_{bb}+\Gamma_{c}\rho_{cc}-\frac{{i}\Omega_{\text{p}}}{2}(\rho_{ab}-\rho_{ba})\nonumber\\
&+\frac{{i}\Omega_{\text{c}}}{2}(\rho_{bc}-\rho_{cb}),\label{eq:EMOTION2}\\
\dot{\rho}_{cc}=&-\Gamma_{c}\rho_{cc}-\frac{{i}\Omega_{\text{c}}}{2}(\rho_{bc}-\rho_{cb}),\label{eq:EMOTION3}\\
\dot{\rho}_{ab}=&-({i}\Delta_{\text{p}}+\frac{\Gamma_{b}}{2}+\gamma_{\rm p})\rho_{ab}\nonumber\\&-\frac{{i}\Omega_{\text{p}}}{2}(\rho_{bb}-\rho_{aa})+\frac{{i}\Omega_{\text{c}}}{2}\rho_{ac},\label{eq:EMOTION4}\\
\dot{\rho}_{bc}=&-({i}\Delta_{\text{c}}-2{i}\Sigma_c\sin^2\omega_{\text{rf}}t+\frac{\Gamma_{b}+\Gamma_{c}}{2}+\gamma_{\rm p}+\gamma_{\rm c})\rho_{bc}\nonumber\\-&\frac{{i}\Omega_{\text{c}}}{2}(\rho_{cc}-\rho_{bb})-\frac{{i}\Omega_{\text{p}}}{2}\rho_{ac},\label{eq:EMOTION5}\\
\dot{\rho}_{ac}=&-({i}\Delta_{\rm p} + {i}\Delta_{\rm c} -2{i}\Sigma_c\sin^2\omega_{\text{rf}}t+\frac{\Gamma_{c}}{2}+\gamma_{\rm c})\rho_{ac}\nonumber\\&+\frac{{i}\Omega_{\text{c}}}{2}\rho_{ab}-\frac{{i}\Omega_{\text{p}}}{2}\rho_{bc}.\label{eq:EMOTION6}
\end{flalign}
\end{subequations}
%%%%%%%%%%%%%%%%%
%%%%%%%%%%%%%%%%%
As in Sec.\ \ref{sec:3}, 
$\Gamma_{c}$ and $\gamma_{\rm c}$ denote the natural width of state $|c\rangle$
and the linewidth of the control laser. $\Gamma_{b}$ and $\gamma_{\rm p}$,
which
were not introduced in Sec.\ \ref{sec:3},
are the corresponding quantities for state $|b\rangle$ and the probe laser
(Fig.\ \ref{fig:Fig1}).
Correspondingly, the
optical Bloch equations in the $N$-level approximation now read
\begin{subequations}
\begin{flalign}
\dot{\rho}_{aa}= & \Gamma_b \rho_{bb} + \frac{{i}\Omega_{\text{p}}}{2}
(\rho_{ab}-\rho_{ba}), \label{eq:HRHON31} \\
\dot{\rho}_{bb}= & -\Gamma_b \rho_{bb} - \frac{{i}\Omega_{\text{p}}}{2}
(\rho_{ab}-\rho_{ba})  \nonumber \\
& + \Gamma_{{c}}\sum_k \rho_{kk}+
\frac{{i}\Omega_{\text{c}}}{2}
\sum_k (J_k\rho_{bk}-J_k\rho_{kb}),\label{eq:HRHON32}\\
\dot{\rho}_{ab}=& -\left({\Gamma_b \over 2}+\gamma_{\rm p}+{
 i}\Delta_{\text{p}}\right)
\rho_{ab}+
\frac{{i}\Omega_{\text{p}}}{2}(\rho_{aa}-\rho_{bb})\nonumber \\
& +\frac{{i}\Omega_{\text{c}}}{2}\sum_k J_k \rho_{ak}
,\label{eq:HRHON33} \\
\dot{\rho}_{ak}=& -\left[{\Gamma_c \over 2}+\gamma_{\rm p}+
\gamma_{\rm c}+{
 i}(\Delta_{\rm p}+\Delta_{\text{c}}-\Sigma_c+2k\omega_{\rm rf})\right]
\rho_{ak}\nonumber\\
& -\frac{{i}\Omega_{\text{p}}}{2}\rho_{bk}
+\frac{{i}\Omega_{\text{c}}J_k}{2}\rho_{ab}
,\label{eq:HRHON34} \\
\dot{\rho}_{bk}=& -\left[{\Gamma_b+\Gamma_c \over  2}+\gamma_{\rm c}+{
 i}(\Delta_{\text{c}}-\Sigma_c+2k\omega_{\rm rf})\right]
\rho_{bk}\nonumber\\
&+\frac{{i}\Omega_{\text{c}}}{2}\left(\rho_{bb}-\sum_j J_j\rho_{jk}\right)
-\frac{{i}\Omega_{\text{p}}}{2}\rho_{ak}
,\label{eq:HRHON35} \\
\dot{\rho}_{jk}=& -[\Gamma_{{c}}-2{i}(j-k)\omega_{\rm rf}]\rho_{jk}
+\frac{{i}\Omega_{\text{c}}}{2}(
J_k\rho_{jb}-J_j\rho_{bk}),\label{eq:HRHON36}
\end{flalign}
\end{subequations} 
with the indexes $j$ and $k$ running over all the components of the Floquet
manifold of state $|c\rangle$.
These equations are to the Hamiltonian 
$\tilde{\sf H}_{3{\rm F}}'$ what Eqs.\ (\ref{eq:EMOTION1}--\ref{eq:EMOTION6})
are to the Hamiltonians ${\sf H}_{3{\rm F}}$ and ${\sf H}_{3{\rm F}}'$.
As in the 2-state case, the solutions of Eqs.\ 
(\ref{eq:HRHON31}--\ref{eq:HRHON36}) tend to constants in the long time limit,
whereas those of Eqs.\
(\ref{eq:EMOTION1}--\ref{eq:EMOTION6}) tend to a steady state in which
the elements of the density matrix oscillate at a fundamental angular frequency
$2 \omega_{\rm rf}$.
In either case,
the absorption coefficient for the probe beam is given by the equation
\begin{equation}
\alpha=- k_{\rm p}\,\mbox{Im}\,\left[
{2\pi {\cal N} |{\bf d}_{ba}\cdot \hat{\epsilon}_{\rm p}|^2 \over
 \epsilon_0\hbar\Omega_{\rm p}}\,\langle \rho_{ba}(\infty)\rangle_{\rm av}\right],
\end{equation}
where $k_{\rm p}$ is the wave number of the probe field, 
${\cal N}$ is the number density of the atoms forming the medium,
${\bf d}_{ba}$ is the matrix element of the dipole operator between the states
$|a\rangle$ and $|b\rangle$, $\hat{\epsilon}_{\rm p}$ is the polarization vector of the probe
field,
$\epsilon_0$ is the permittivity of free space,
and $\langle \rho_{ba}(\infty)\rangle_{\rm av}$ denotes the average
value of $\rho_{ba}$ in the stationary regime.
For a cold atomic ensemble, $\langle \rho_{ba}(\infty)\rangle_{\rm av}\equiv
\langle \rho_{ba}(\infty)\rangle$, with
\begin{equation}\label{eq:AV2}
\langle\rho_{ba}(\infty)\rangle=\frac{1}{T}\lim_{t\to\infty}\int_{t}^{t+T}
{\rho_{ba}(t')\,`{d}t'}.
\end{equation}
For a thermal ensemble where the Doppler effect must be taken into account,
$\langle \rho_{ba}(\infty)\rangle_{\rm av}$ is
the average of $\langle\rho_{ba}(\infty)\rangle$
over the Maxwellian distribution of velocity of the atoms.

%%%%%%%%%%%%%%%%%%%%%%%%%%%%%%%%%%%%%%%%%%%%%%%%%%%%%%%%%%%
\subsection{Absorption sidebands and induced sidebands}\label{subsec:4b}
\subsubsection{Weak control fields}\label{subsubsec:4b1}

%%%%%%%%%%%%%%%%%%%%%%%%%%%%%%%%%%%%%%%%%%%%%%%%
%%%%%%%%%%%%%%%%%%%%%%%%%%%%%%%%%%%%%%%%%%%%%%%%
%%%%%%%%%%%%%%%%%%%%%%%%%%%%%%%%%%%%%%%%%%%%%%%%
\begin{figure}[!t]
\begin{center}
\includegraphics*[width=8.5cm]{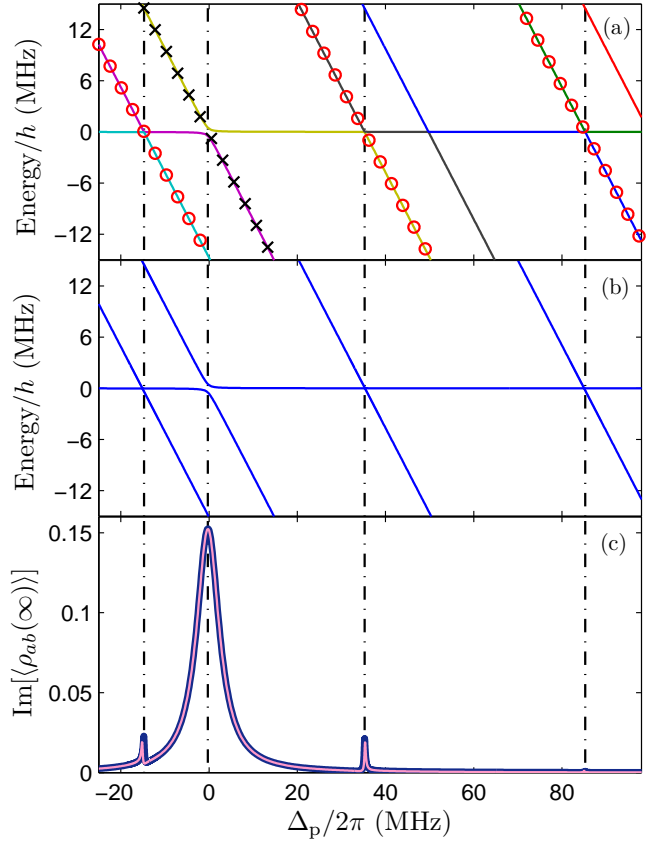}
\caption{\label{fig:Fig5}
(color online)
The quasienergy spectrum, (a) and (b), and the absorption profile of the
probe field, (c), in an rf-dressed 3-level system.
$\Omega_{\text{p}}/2\pi=1$ MHz, $\Omega_{\text{c}}/2\pi=8$ MHz, 
$\Sigma_c/2\pi=35$ MHz, $\Delta_{\text{c}}/2\pi=0$ MHz, 
$\omega_{\text{rf}}/2\pi=25$ MHz, $\Gamma_b/2\pi=6$ MHz,
$\Gamma_c/2\pi=0.01$ MHz
and $\gamma_{\rm p}/2\pi=\gamma_{\rm c}/2\pi=0.1$ MHz.
(a) and thick dark blue curve in (c): exact results.
(b) and thin light pink curve in (c): results in the $N$-level approximation.
In (a), the crosses identify $\epsilon_{b,0}$ and the circles
$\epsilon_{c,0}$, $\epsilon_{c,1}$ and $\epsilon_{c,-1}$.
The other oblique lines correspond to $\epsilon_{b,n}$, $n\not=0$.
The horizontal line
corresponds to $\epsilon_{a,0}$.
One may note small differences in the position of the crossings between
(a) and (b). Because of the significant Stark shift of the upper state,
the control laser is off resonance although $\Delta_{\text{c}}/2\pi=0$, which
explains why the absorption profile of the probe field does not exhibit EIT
dips.
}
\end{center}
\end{figure}
%%%%%%%%%%%%%%%%%%%%%%%%%%%%%%%%%%%%%%%%%%%%%%%%
%%%%%%%%%%%%%%%%%%%%%%%%%%%%%%%%%%%%%%%%%%%%%%%%

A representative section of the quasienergy spectrum
of the full Floquet Hamiltonian is shown in 
{{\color{black}Fig.\ \ref{fig:Fig5}(a)}} for a case where the control laser
is sufficiently weak that the inequality (\ref{eq:COND1}) is fulfilled. 
The horizontal line interrupted by avoided crossings
indicates the quasienergy of the zeroth order Floquet
sideband of state $|a\rangle$. The oblique lines arise either
from the sidebands of
the state $|b\rangle$ or from those of the state $|c\rangle$, as indicated in the caption. 
The $b$- and $c$-quasienergy manifolds run parallel to each other in
{{\color{black}Fig.\ \ref{fig:Fig5}(a)}}, although they intersect
in {{\color{black}Fig.\ \ref{fig:Fig3}(a)}}, 
because the parameter varied in {{\color{black}Fig.\ \ref{fig:Fig5}}}
is the detuning of the probe field while that varied in
{{\color{black}Fig.\ \ref{fig:Fig3}}} is the detuning of the control field.
The latter is actually set to zero in these calculations,
which means, in view of the Stark shift of state $|c\rangle$,
that the control laser is off resonance.
The quasienergy spectrum obtained in the $N$-level approximation is shown
in {{\color{black}Fig.\ \ref{fig:Fig5}(b)}}, for comparison. The
absorption profile for a cold atom ensemble
is given in {{\color{black}Fig.\ \ref{fig:Fig5}(c)}}. (The quantity represented
in
{{\color{black}Fig.\ \ref{fig:Fig5}(c)}} is $\langle \rho_{ba}(\infty)\rangle$, 
which is proportional to the absorption coefficient if the atomic motion can be neglected. The maxima of  $\langle \rho_{ba}(\infty)\rangle$ correspond to maxima of absorption.)

The dressed states crossing the horizontal lines all contain
the bare state $|b\rangle$ --- i.e., at least one of the 
$c_{b,n}$ coefficients is non-zero in Eq.\ (\ref{eq:newpsi}).
Therefore each of these crossings
corresponds to a resonant coupling between this state
and the bare state $|a\rangle$ and may give rise 
to an enhancement of absorption of the probe field coupling
$|a\rangle$ to $|b\rangle$.
Fig.\ \ref{fig:Fig5}(c) shows a
large enhancement of absorption
at the crossing involving the zeroth order sideband of
state $|b\rangle$ (at $\Delta_{\rm p} = 0$), and smaller enhancements at the
other crossings. The enhancement at the crossings with the $+1$ sideband
of $|c\rangle$ (at 85 MHz) is hardly visible, though,
and the enhancements associated with the crossings of the $+1$ and $+2$ 
sidebands of $|b\rangle$ (at 50 and 100 MHz, respectively) are too small to be seen.
These enhancements are weaker compared to those involving the 
zeroth-order sidebands both because of the scaling of $\Omega_{\rm c}$ by
the $J_n$ factors
($|J_n|$ is much smaller than 1 when $n\not= 0$ for the parameters 
of {{\color{black}Fig.\ \ref{fig:Fig5}}}) and because 
of the larger energy difference between the respective sidebands
(a larger energy difference means a smaller 
admixture of $|b\rangle$ into the sidebands of $|c\rangle$).
However, as will be seen shortly, significant enhancements may occur at 
the crossings with the $n\not=0$ sidebands of $b$ for stronger
control fields.

Comparing 
 {{\color{black}Fig.\ \ref{fig:Fig5}(a)}} to
 {{\color{black}Fig.\ \ref{fig:Fig5}(b)}} and 
the thick blue line to the thin pink line in
{{\color{black}Fig.\ \ref{fig:Fig5}(c)}}, we see that the
$N$-level approximation holds well in the present case, as could have
been expected since the condition (\ref{eq:MODEL}) is fulfilled.
Because the side bands of the $b$-state are neglected
in the $N$-level approximation,
the $b$-quasienergy manifold of {{\color{black}Fig.\ \ref{fig:Fig5}(a)}}
reduces to a single quasienergy curve
in {{\color{black}Fig.\ \ref{fig:Fig5}(b)}}. However,
the $c$-manifold is practically the
same in the two spectra. The absorption profile is also very well reproduced
in the $N$-level approximation for the parameters considered: the two
sets of results are almost identical apart for very
small differences in the position of the
enhancements (noticeable only for the enhancement at $\Delta_{\rm p}=-15$ MHz
on the scale of the figure) and very small differences in their amplitude.

%%%%%%%%%%%%%%%%%%%%%%%%%%%%%%%%%%%%%%%%%%%%%%%%%%%%%%%%%%%%%%%%%%%%%%%%%%%%%%%%%%%%%%%%%%%%%%%%%%%%%%%%%%%%%%%%%%%%%%%%%%%%%%%%%%%%%%%%%%%%%%%%%%%%%%%%%%%%%%%%%%%%%%%%%%%%%%%%%%%%%%%%%%%%%%%%%%%%%%%%%%%%%%%%%%%%%%%%%%%%%%%%%%%%%%%%%%%%%%%%%%%%%%%%%%%%%%%%%%%%%%%%%%%%%%%%%%%%%%%%%%%%%%%%%%%%%%%%%%%%%%%%
%%%%%%%%%%%%%%%%%%%%%%%%%%%%%%%%%%%%%%%%%%%%%%%%%%%%%%%%%%%%%%%%%%%%%%%%%%%%%%%%%%%%%%%%%%%%%%%%%%%%%%%%%%%%%%%%%%%%%%%%%%%%%%%%%%%%%%%%%%%%%%%%%%%%%%%%%%%%%%%%%%%%%%%%%%%%%%%%%%%%%%%%%%%%%%%%%%%%%%%%%%%%%%%%%%%%%%%%%%%%%%%%%%%%%%%%%%%%%%%%%%%%%%%%%%%%%%%%%%%%%%%%%%%%%%%%%%%%%%%%%%%%%%%%%%%%%%%%%%%%%%%%
\subsubsection{Strong control fields}\label{subsubsec:4b2}

\begin{figure}[!t]
\begin{center}
\includegraphics*[width=8.5cm]{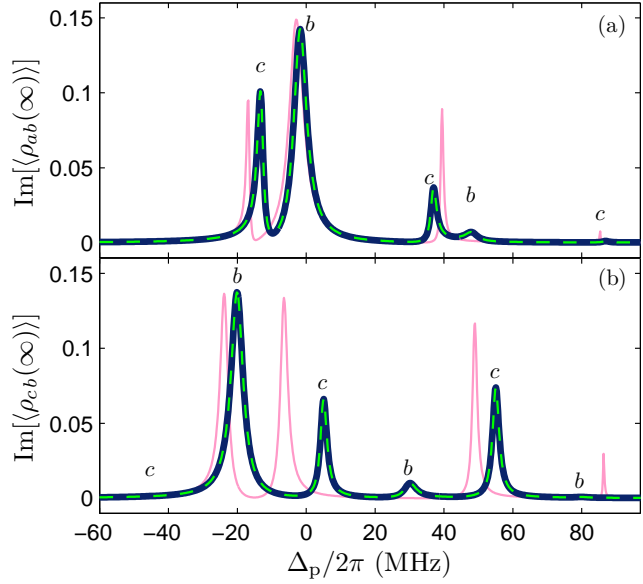}
\caption{\label{fig:Fig6}
(color online)
The probe absorption in the same rf-dressed 3-level
system as in Fig.\ \ref{fig:Fig5} but for stronger
control fields for which the $N$-level approximation is inaccurate.
$\Omega_{\text{c}}/2\pi=30$ MHz in (a) and 60 MHz in (b).
Thick blue solid curves: results obtained by solving the full
optical Bloch equations,
Eqs.\ (\ref{eq:EMOTION1}--\ref{eq:EMOTION6}).
Thin pink solid curves: results obtained in the $N$-level approximation,
Eqs.\ (\ref{eq:HRHON31}--\ref{eq:HRHON36}).
Dashed green curves: results obtained by adding the sidebands of 
state $|b\rangle$, but not those of state $|a\rangle$, to the $N$-level approximation.
The enhancements associated with crossings
between the lower state and the intermediate state manifold are identified
by the letter $b$
between the lower state and the
Rydberg state manifold by the letter $c$.
}
\end{center}
\end{figure}

We now increase $\Omega_{\rm c}/2\pi$ 
to 30 and 60 MHz, keeping the other parameters the same
as in Fig.\ \ref{fig:Fig5}.
The resulting absorption profiles are shown in
Figs.\ \ref{fig:Fig6}(a) and  \ref{fig:Fig6}(b), respectively. 
As expected from the previous discussion, the $N$-level approximation
deteriorates as the strength of the control field increases. It is still
in rough agreement with the exact result 
in the case of Fig.\ \ref{fig:Fig6}(a),
for which $\Omega_{\rm c}$ is smaller, but not much smaller,
than $2 \omega_{\rm rf}$,
although there are differences in the position and the amplitudes of the
enhancements. However, when the condition (\ref{eq:MODEL}) is more strongly 
violated, the $N$-level approximation breaks down completely.
There is little agreement with the exact result in 
Fig.\ \ref{fig:Fig6}(b).

An interesting feature of Fig.\ \ref{fig:Fig6}(a) is the presence of a
small enhancement of absorption at $\Delta_{\rm p}/2\pi \approx 48$ MHz.
This enhancement is not present in the $N$-level approximation. It coincides
with the crossing of $|a\rangle$ with the $+1$ sideband
of $|b\rangle$ in the quasienergy spectrum, at $2\hbar\omega_{\rm rf}$ above the
main absorption peak (which is concomitant with the crossing of the
0-th order sideband of $|b\rangle$.) 
This feature is more prominent in Fig.\ \ref{fig:Fig6}(b), at
$\Delta_{\rm p}/2\pi \approx 30$ MHz;
the $+2$ sideband of $|b\rangle$
is also (barely) visible, at about 80 MHz at this higher power of the
control laser.
We refer to such sidebands as ``induced sidebands", as they arise from
sidebands of $|b\rangle$ induced by the coupling of this state with
the sidebands of $|c\rangle$ by the control field.
(Recall that the rf field is assumed to be too weak to
dress state $|b\rangle$ directly: the $b$-manifold originates from the coupling of
$|b\rangle$ to the $c$-manifold by the control field, not from the coupling of $|b\rangle$ to itself by the rf field.)

That the sidebands of $|a\rangle$ play no role in this absorption profile
is shown by the good agreement
between the exact results (thick blue curves)
and the results obtained when these sidebands are neglected 
altogether (green dashed curves).

Before closing, we briefly comment on whether the conditions
{{\color{black}Eq.\ (\ref{eq:COND1}) 
and (\ref{eq:COND2})}} are sufficient to garantee the validity of
the $N$-level approximation for a three-state system.
Recall that these conditions mean 
that the oscillations of the rf field are faster than the time scales
over which the system evolves under the effect of the control laser field
and the spontaneous
decay of state $|c\rangle$. 
However, the probe field introduces an additional time
scale in the problem: if this field is excessively strong, it will
disturb the dynamics of the $b$-$c$ system too rapidly for the state $|c\rangle$ to
show a manifold structure. 
In this case, the
Floquet sidebands of state $|a\rangle$ induced by its coupling to state $|c\rangle$
can no longer be ignored.
Given that the effective Rabi frequency for the coupling of $|a\rangle$ to $|c\rangle$
is approximately given by
$\Omega_{\text{p}}\Omega_{\text{c}}/2\Sigma_c$
{{\color{black}\cite{Radmore,linskens}}}, we expect that 
the condition 
%%%%%%%%%%%%%%%
%%%%%%%%%%%%%%% 
\begin{equation}\label{eq:COND3}
|\frac{\Omega_{\text{p}}\Omega_{{\text{c}}}}{ 2\Sigma_c}|\ll2\omega_{\text{rf}}
\end{equation}
%%%%%%%%%%%%%%%
%%%%%%%%%%%%%%% 
should be added to the conditions 
{{\color{black}Eq.\ (\ref{eq:COND1}) 
and (\ref{eq:COND2})}} for the $N$-level approximation to be valid.
Should the inequality (\ref{eq:COND3}) not be fulfilled, then it is likely
that the Floquet sidebands of $|a\rangle$ need to be taken into account,
as well as those of $|b\rangle$ and $|c\rangle$.

%%%%%%%%%%%%%%%%%%%%%%%%%%%%%%%%%%%%%%%%%%%%%%%%
%%%%%%%%%%%%%%%%%%%%%%%%%%%%%%%%%%%%%%%%%%%%%%%%
%%%%%%%%%%%%%%%%%%%%%%%%%%%%%%%%%%%%%%%%%%%%%%%%
%%%%%%%%%%%%%%%%%%%%%%%%%%%%%%%%%%%%%%%%%%%%%%%%
%%%%%%%%%%%%%%%%%%%%%%%%%%%%%%%%%%%%%%%%%%%%%%%%
%%%%%%%%%%%%%%%%%%%%%%%%%%%%%%%%%%%%%%%%%%%%%%%%
%%%%%%%%%%%%%%%%%%%%%%%%%%%%%%%%%%%%%%%%%%%%%%%%
%%%%%%%%%%%%%%%%%%%%%%%%%%%%%%%%%%%%%%%%%%%%%%%%
%%%%%%%%%%%%%%%%%%%%%%%%%%%%%%%%%%%%%%%%%%%%%%%%%%%%%%%%%%%%%%%%%%%%%%%%%%%%%%%%%%%%%%%%%%%%%%%%%%%%%%%%%%%%%%%%%%%%%%%%%%%%%%%%%%%%%%%%%%%%%%%%%%%%%%%%%%%%%%%%%%%%%%%%%%%%%%%%%%%%%%%%%%%%%%%%%%%%%%%%%%%%%%%%%%%%%%%%%%%%%%%%%%%%%%%%%%%%%%%%%%%%%%%%%%%%%%%%%%%%%%%%%%%%%%%%%%%%%%%%%%%%%%%%%%%%%%%%%%%%%%%%
%%%%%%%%%%%%%%%%%%%%%%%%%%%%%%%%%%%%%%%%%%%%%%%%%%%%%%%%%%%%%%%%%%%%%%%%%%%%%%%%%%%%%%%%%%%%%%%%%%%%%%%%%%%%%%%%%%%%%%%%%%%%%%%%%%%%%%%%%%%%%%%%%%%%%%%%%%%%%%%%%%%%%%%%%%%%%%%%%%%%%%%%%%%%%%%%%%%%%%%%%%%%%%%%%%%%%%%%%%%%%%%%%%%%%%%%%%%%%%%%%%%%%%%%%%%%%%%%%%%%%%%%%%%%%%%%%%%%%%%%%%%%%%%%%%%%%%%%%%%%%%%%
\subsection{EIT in rf-dressed atomic ensembles}\label{subsec:4d}
\subsubsection{Cold atomic ensembles}\label{subsubsec:4d1}
%%%%%%%%%%%%%%%%%%%%%%%%%%%%%%%%%%%%%%%%%%%%%%%%
%%%%%%%%%%%%%%%%%%%%%%%%%%%%%%%%%%%%%%%%%%%%%%%%
%%%%%%%%%%%%%%%%%%%%%%%%%%%%%%%%%%%%%%%%%%%%%%%%
%%%%%%%%%%%%%%%%%%%%%%%%%%%%%%%%%%%%%%%%%%%%%%%%
\begin{figure}[!t]
\begin{center}
\includegraphics*[width=8.5cm]{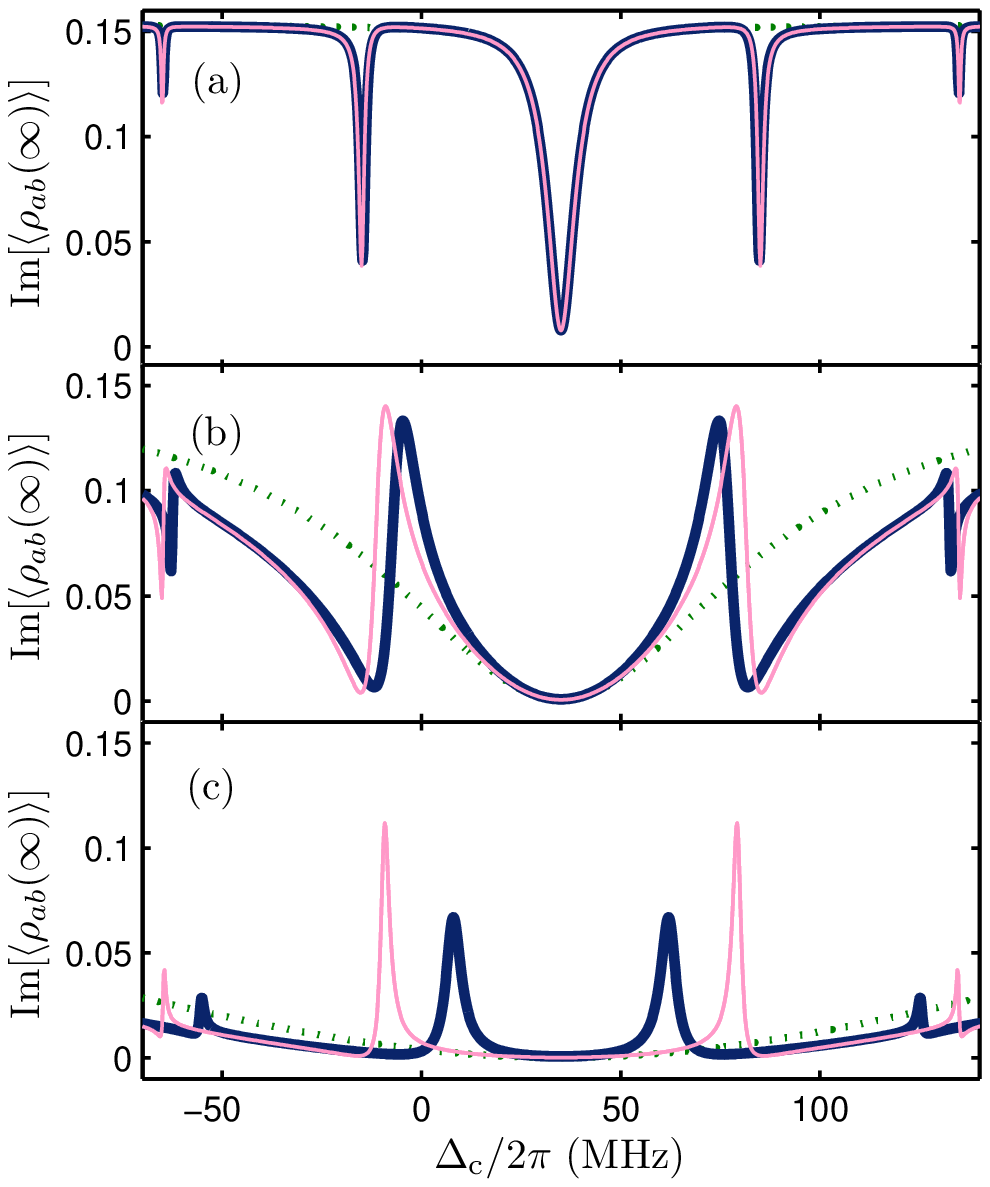}
\caption{\label{fig:Fig7}
(color online)
Probe absorption vs.\ control laser detuning in an rf-dressed 3-level system,
for three different control
Rabi frequencies. The parameters of the system are the same as
in Figs.\ \ref{fig:Fig5} and \ref{fig:Fig6}, but here the probe field is
on-resonance
($\Delta_{\rm p}=0$) and $\Delta_{\rm c}$ varies.
$\Omega_{\rm c}/2\pi$ is 8 MHz in (a), 30 MHz in (b) and 60 MHz in (c).
Thick blue solid curves: results obtained by solving the full
optical Bloch equations,
Eqs.\ (\ref{eq:EMOTION1}--\ref{eq:EMOTION6}).
Thin pink solid curves: results obtained in the $N$-level approximation,
Eqs.\ (\ref{eq:HRHON31}--\ref{eq:HRHON36}).
Dotted green curves: absorption profile without rf field (the energy
of the Rydberg state was shifted by $\Sigma_c$ to facilitate the comparison
with the other results).
As $\Omega_{\rm c}$ increases, the EIT transparency window in (a) expands
into the sidebands region and
the EIT absorption dips change into absorption peaks.
The width
of the features is modulated by the square of the Bessel function factors
$J_n$.
}
\end{center}
\end{figure}
%%%%%%%%%%%%%%%%%%%%%%%%%%%%%%%%%%%%%%%%%%%%%%%%
%%%%%%%%%%%%%%%%%%%%%%%%%%%%%%%%%%%%%%%%%%%%%%%%
%%%%%%%%%%%%%%%%%%%%%%%%%%%%%%%%%%%%%%%%%%%%%%%%
%%%%%%%%%%%%%%%%%%%%%%%%%%%%%%%%%%%%%%%%%%%%%%%%
We still assume that the Doppler effect is negligible, as in the last section.
However,
we now set $\Delta_{\rm p}$ to 0 and vary $\Delta_{\rm c}$ rather than set
$\Delta_{\rm c}$ to 0 and vary $\Delta_{\rm p}$. If the rf field
was absent, the control laser field would couple the states $|b\rangle$ and $|c\rangle$
resonantly at $\Delta_{\rm c}=0$, which would be accompanied by a dip
in Im$[\langle\rho_{ab}(\infty)\rangle]$). (This is the well known
EIT absorption window at $\Delta_{\rm p}=0$ \cite{eit_review}.) In the presence of the rf field,
this dip is displaced to $\Delta_{\rm c} = \Sigma_c$, due to the ac Stark shift
of $|c\rangle$, and splits into multiple sidebands.

EIT with rf modulation is illustrated by
Fig.\ \ref{fig:Fig7}. These results were calculated for
the same values of $\Omega_{\rm p}$,
$\Sigma_c$, $\omega_{\rm rf}$, $\Gamma_{b}$, $\Gamma_{c}$, $\gamma_{\rm p}$
and $\gamma_{\rm c}$ as in Figs.\ \ref{fig:Fig4} and  \ref{fig:Fig5}. The
probe Rabi frequency, $\Omega_{\rm p}/2\pi$, increases from 8 MHz in
Fig.\ \ref{fig:Fig7}(a) to 30 MHz in
Fig.\ \ref{fig:Fig7}(b) and to 60 MHz in
Fig.\ \ref{fig:Fig7}(c).
As seen from the figure,
the rf field creates transmission sidebands on either side of the main
EIT feature. Since $\Delta_{\rm p}$ is set to zero and the 
frequency of the control field is varied, these sidebands occur at the 
values of $\Delta_{\rm c}$ at which
the $c$-quasienergy manifold intersects the zeroth-order quasienergy sideband
of the state $|b\rangle$. In the weak control case of
Fig.\ \ref{fig:Fig7}(a), they manifest as narrow dips regularly spaced
by $2\omega_{\rm rf}$ \cite{newnote}.
Apart for minor differences in the depth,
the $N$-state approximation reproduces these features very well (the thick
blue curve to the thin pink curve).
The absorption profile is thus well explained by the
model in which the upper state simply splits into a comb of Floquet states,
each one interacting with the intermediate state
independently from the others.
However, this picture changes when $\Omega_{\rm c}$ approaches
or exceeds $2\omega_{\rm rf}$ --- see, respectively, Fig.\ \ref{fig:Fig7}(b)
and (c): the EIT dips broaden and shift
towards the zeroth-order sideband when $\Omega_{\rm c}$ increases, and the
absorption profile first acquires a peak-and-trough structure and then
changes into a series of absorption peaks (rather
than absorption dips) superimposed on a slowly varying background.
These peaks occur at the detunings at which state $|a\rangle$ is resonant with
the dressed state formed by the states $|b\rangle$ and $|c\rangle$ coupled
by the control field.
The same changes are observed in
the $N$-level approximation, at least for the parameters considered, but
the position of the absorption features is more and more incorrect.

The green dotted curves represent the absorption profile calculated without the
sidebands of state $|c\rangle$ (but taking its Stark shift into account).
Comparing these results to the thick blue curve in
Fig.\ \ref{fig:Fig7}(c) shows that the suppression of EIT in narrow ranges
of frequencies at large control powers arises from the interplay
between these sidebands.

%%%%%%%%%%%%%%%
%%%%%%%%%%%%%%% 
%%\begin{equation}\label{eq:DARKSHIFT}
%%\delta_{n}=\sum_{m\neq -n}{(\Omega_{\text{c}}J_{m}/2)^2\over2\omega_{\text{rf}}(m+n)}.
%%\end{equation}
%%%%%%%%%%%%%%%
%%%%%%%%%%%%%%% 
%%Then the resonance position is
%%$\Delta_{\text{c}}^{n}=\Sigma_c\pm2n\omega_{\text{rf}}\mp\delta_{n}$.
%%However, the $N$-level approximation gives
%%the resonance positions at
%%$\Delta_{\text{c}}^{n}=\Sigma_c\pm2n\omega_{\text{rf}}$.
%%This is because there are no perturbation
%%to $\ket{c',n}$ from the sidebands
%%of the intermediate state.
%%%%%%%%%%%%%%%%%%%%%%%%%%%%%%%%%%%%%%%%%%%%%%%%%%%%%%%%%%%%%%%%%%%%%%%%%%%%%%%%%%%%%%%%%%%%%%%%%%%%%%%%%%%%%%%%%%%%%%%%%%%%%%%%%%%%%%%%%%%%%%%%%%%%%%%%%%%%%%%%%%%%%%%%%%%%%%%%%%%%%%%%%%%%%%%%%%%%%%%%%%%%%%%%%%%%%%%%%%%%%%%%%%%%%%%%%%%%%%%%%%%%%%%%%%%%%%%%%%%%%%%%%%%%%%%%%%%%%%%%%%%%%%%%%%%%%%%%%%%%%%%%
%%%%%%%%%%%%%%%%%%%%%%%%%%%%%%%%%%%%%%%%%%%%%%%%%%%%%%%%%%%%%%%%%%%%%%%%%%%%%%%%%%%%%%%%%%%%%%%%%%%%%%%%%%%%%%%%%%%%%%%%%%%%%%%%%%%%%%%%%%%%%%%%%%%%%%%%%%%%%%%%%%%%%%%%%%%%%%%%%%%%%%%%%%%%%%%%%%%%%%%%%%%%%%%%%%%%%%%%%%%%%%%%%%%%%%%%%%%%%%%%%%%%%%%%%%%%%%%%%%%%%%%%%%%%%%%%%%%%%%%%%%%%%%%%%%%%%%%%%%%%%%%%
\subsubsection{Thermal atomic ensembles}\label{subsubsec:4d2}
%%%%%%%%%%%%%%%%%%%%%%%%%%%%%%%%%%%%%%%%%%%%%%%%
%%%%%%%%%%%%%%%%%%%%%%%%%%%%%%%%%%%%%%%%%%%%%%%%
%%%%%%%%%%%%%%%%%%%%%%%%%%%%%%%%%%%%%%%%%%%%%%%%
%%%%%%%%%%%%%%%%%%%%%%%%%%%%%%%%%%%%%%%%%%%%%%%%
\begin{figure}[!t]
\begin{center}
\includegraphics*[width=8.5cm]{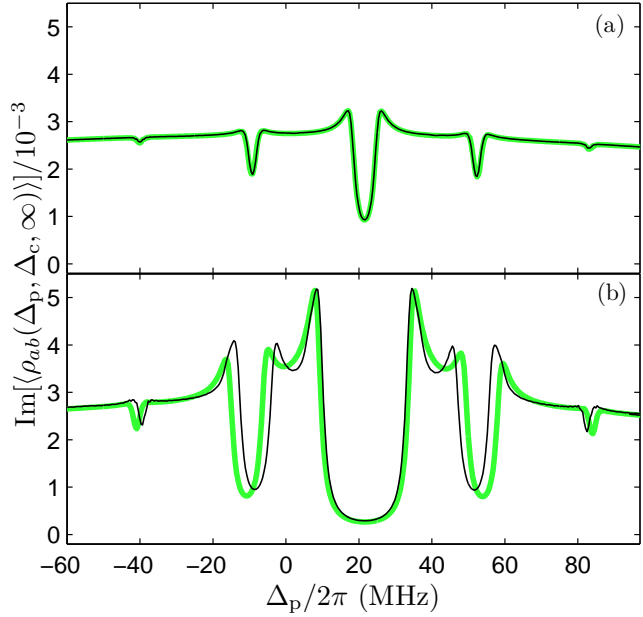}
\caption{\label{fig:Fig8}
(color online)
The probe absorption in the same rf-dressed 3-level
system as in Fig.\ \ref{fig:Fig7} but here with Doppler
averaging and plotted against probe detuning.
The probe and control laser beams are assumed to be colinear and
counterpropagating.
$\Omega_{\rm c}/2\pi$ is 8 MHz in (a) and 30 MHz in (b).
$\Delta_{\rm c}=0$ in both graphs.
Thick green curves: results obtained by solving the full
optical Bloch equations,
Eqs.\ (\ref{eq:EMOTION1}--\ref{eq:EMOTION6}).
Thin black curves: results obtained in the $N$-level approximation,
Eqs.\ (\ref{eq:HRHON31}--\ref{eq:HRHON36}).
The higher power of the control field in (b) results in a broadening
and a shift of the EIT
dips and to inaccuracies in the $N$-level approximation. 
}
\end{center}
\end{figure}
%%%%%%%%%%%%%%%%%%%%%%%%%%%%%%%%%%%%%%%%%%%%%%%%
%%%%%%%%%%%%%%%%%%%%%%%%%%%%%%%%%%%%%%%%%%%%%%%%
%%%%%%%%%%%%%%%%%%%%%%%%%%%%%%%%%%%%%%%%%%%%%%%%
%%%%%%%%%%%%%%%%%%%%%%%%%%%%%%%%%%%%%%%%%%%%%%%%
Let us assume that the two laser beams are colinear and 
counterpropagating.
Compared to an atom at rest, the probe and control angular frequencies for
an atom moving with a velocity component $v$ in the direction of the control
laser beam are
shifted respectively by $k_{\text{p}}v$ and $-k_{\text{c}}v$,
where $k_{\text{p}}$ and 
$k_{\text{c}}$ are the wave numbers in the laboratory frame.
In the reference frame of the atom,
the probe detuning is therefore 
upshifted to $\Delta_{\text{p}}+k_{\text{p}}v$
and the control detuning downshifted to
$\Delta_{\text{c}}-k_{\text{c}}v$, where 
$\Delta_{\text{p}}$ and $\Delta_{\text{c}}$ are the detunings
in the laboratory frame.
The absorption of the probe beam at 
$\Delta_{\text{p}}$ and $\Delta_{\text{c}}$
will thus be determined by the velocity average 
of the steady-state coherence $\langle\rho_{ab}(\infty)\rangle$ 
calculated at the Doppler-shifted detunings.
We denote this average by
$\langle\rho_{ij}(\Delta_{\text{p}},\Delta_{\text{c}},\infty)\rangle_{\rm av}$.
For atoms of mass $m$ at a temperature $T$,
%%%%%%%%%%%%%%%
%%%%%%%%%%%%%%% 
\begin{equation}
\langle\rho_{ij}(\Delta_{\text{p}},\Delta_{\text{c}},\infty)\rangle_{\rm av}=
\int_{-\infty}^{\infty}{f(v)\langle\rho_{ij}(\Delta_{\text{p}}+k_{\text{p}}v,\Delta_{\text{c}}-k_{\text{c}}v,\infty)\rangle\,{d}v}
\end{equation}
%%%%%%%%%%%%%%%
%%%%%%%%%%%%%%% 
with
\begin{equation}
f(v)=\sqrt{m\over 2\pi k{_{\rm B} T}}
 \exp\left(-{m v^2 \over 2 k_{\rm B} T}\right),
\end{equation}
where $k_{\rm B}$ is the Boltzmann constant.

{{\color{black}Fig.\ \ref{fig:Fig8}}} shows the probe absorption calculated
using the $N$-level approximation (thin black curves) or calculated
directly from 
{{\color{black}Eqs.\ (\ref{eq:EMOTION1}--\ref{eq:EMOTION6})}} (thick green
curves),
as a function of the probe detuning,
for two different values of $\Omega_{\text{c}}$. The other parameters
are the same as those used in {{\color{black}Fig.\ \ref{fig:Fig5}}}. In
the Doppler average, we assume 
a cloud of $^{85}$Rb atoms at a temperature 
of 40 $^{\rm o}$C, and probe and control wavelengths of
780 and 480 nm, respectively.

We see that the $N$-level approximation is accurate in the case of
figure (a), for which the control power is relatively low.
For small values of
$\Omega_{\text{c}}$,
each absorption minimum corresponds to EIT in
%%The frequency separation between any adjacent
%%EIT resonances is not $2\omega_{\text{rf}}$ as the 
%%distance is scaled by a factor of $k_{\text{p}}/k_{\text{c}}$
%%due to the Doppler averaging. To understand this origin, 
%%we need to use that fact that each EIT resonance corresponds
the velocity class which simultaneously satisfies the two conditions
$\Delta_{\text{p}}+k_{\text{p}}v=0$ 
and $\Delta_{\text{c}}-k_{\text{c}}v-\Sigma_c+2n\omega_{\text{rf}}=0$.
For a fixed value of $\Delta_{\text{c}}$, the corresponding absorption dips
occur at
the probe detunings
%%%%%%%%%%%%%%%
%%%%%%%%%%%%%%% 
\begin{equation}\label{eq:DOPPLER}
\Delta_{\text{p}}=\frac{k_{\text{p}}}{k_{\text{c}}}(\Sigma_c-\Delta_{\text{c}}-2n\omega_{\text{rf}}).
\end{equation}
%%%%%%%%%%%%%%%
%%%%%%%%%%%%%%% 
Adjacent EIT dips are thus separated in probe detuning by
$2\omega_{\text{rf}}\,k_{\text{p}}/k_{\text{c}}$ in these conditions.

Increasing $\Omega_{\rm c}$ both widens the EIP dips through power broadening
and shifts their positions. The latter effect is due to
the shift in the sidebands of $|c\rangle$ arising 
from their coupling with state $|b\rangle$. When this shift is non-negligible,
the second
of the above resonance conditions must be replaced by
$\Delta_{\text{c}}-k_{\text{c}}v-\Sigma_c+2n\omega_{\text{rf}}-\delta_n=0$,
where $\delta_n$ is an 
$\Omega_{\rm c}$-dependent shift, and the absorption dips
occur at
the probe detunings
%%%%%%%%%%%%%%%
%%%%%%%%%%%%%%% 
\begin{equation}\label{eq:DOPPLER2}
\Delta_{\text{p}}=\frac{k_{\text{p}}}{k_{\text{c}}}(\Sigma_c-\Delta_{\text{c}}-2n\omega_{\text{rf}}+\delta_n).
\end{equation}
%%%%%%%%%%%%%%%
As seen from the figure, the $N$-level approximation becomes inaccurate when
$\Omega_{\rm c}/(2 \pi)$ is increased from 8 to 30 MHz: at the higher control
power
there is a clear difference both in the position and in the depth
of the EIT dips between the absorption profiles calculated with and without
making this approximation. (Correcting the $N$-level approximation for the
sidebands of the intermediate state, still neglecting the sidebands of the
lower states, restores the agreement with the exact results.)
%%%%%%%%%%%%%%%%%%%%%%%%%%%%%%%%%%%%%%%%%%%%%%%%%%%%%%%%%%%%%%%%%%%%%%%%%%%%%%%%%%%%%%%%%%%%%%%%%%%%%%%%%%%%%%%%%%%%%%%%%%%%%%%%%%%%%%%%%%%%%%%%%%%%%%%%%%%%%%%%%%%%%%%%%%%%%%%%%%%%%%%%%%%%%%%%%%%%%%%%%%%%%%%%%%%%%%%%%%%%%%%%%%%%%%%%%%%%%%%%%%%%%%%%%%%%%%%%%%%%%%%%%%%%%%%%%%%%%%%%%%%%%%%%%%%%%%%%%%%%%%%%
%%%%%%%%%%%%%%%%%%%%%%%%%%%%%%%%%%%%%%%%%%%%%%%%%%%%%%%%%%%%%%%%%%%%%%%%%%%%%%%%%%%%%%%%%%%%%%%%%%%%%%%%%%%%%%%%%%%%%%%%%%%%%%%%%%%%%%%%%%%%%%%%%%%%%%%%%%%%%%%%%%%%%%%%%%%%%%%%%%%%%%%%%%%%%%%%%%%%%%%%%%%%%%%%%%%%%%%%%%%%%%%%%%%%%%%%%%%%%%%%%%%%%%%%%%%%%%%%%%%%%%%%%%%%%%%%%%%%%%%%%%%%%%%%%%%%%%%%%%%%%%%%
\section{Summary}\label{sec:5}
To conclude, we have investigated the absorption spectrum of
two- and three-level model atomic systems with ac-modulated
quadratic Stark shift, both
in the weak coupling regime and beyond this regime.
Although we specifically consider the case of a
Rydberg system modulated by 
a radio-frequency field, our results are generic.

The different regimes investigated can be defined in terms of the modulation
amplitude of the upper state energy, which is characterized by the quantity
$\Sigma_c$ defined by Eq.\ (\ref{eq:StarkSigma}), and in terms
of the angular frequency
of the rf field,
$\omega_{\rm rf}$, of the Rabi frequency of the control field,
$\Omega_{\rm c}$, and of the natural width of the upper state, $\Gamma_c$.
When the conditions $|\Sigma_c| \ll  2\omega_{\rm rf}$,
$|\Omega_{\rm c}J_0(\Sigma_c/2\omega_{\rm rf})| \ll  2\omega_{\rm rf}$
and $\Gamma_c \ll 2\omega_{\rm rf}$
are all met,
the quasienergy curve crossings in the Floquet spectrum (and the associated
structures in the absorption spectrum) can be treated in isolation of each
other. The system
then behaves as if the most polarizable state is effectively a manifold
of sideband states and the approximation we refer to as the
$N$-level approximation is accurate. This approximation amounts to neglecting
the manifold structure of the less polarizable states and model the ac
modulated system as a many-level system in which these states interact with
a manifold of independent sideband states spawned by the most polarizable
state.

 Increasing the
strength of the control field to
$|\Omega_{\rm c}J_0(\Sigma_c/2\omega_{\rm rf})| \approx 2\omega_{\rm rf}$
shifts the position of the absorption sidebands and changes their
amplitude, in agreement with the perturbative analysis developed in a
different context in Refs.\
\cite{chu} and \cite{hausinger} and with the iterative approach
outlined in Sec.\ \ref{subsubsec:3a2}. The $N$-level approximation becomes
inaccurate as $|\Omega_{\rm c}|$ increases, in particular in predicting
sidebands shifts half too small. However, this approximation
can be brought into agreement with the correct positions of the sidebands
by a simple change in the optical Bloch equations (or the corresponding
Hamiltonian).

In the strong coupling
regime,
where $|\Omega_{\rm c}J_0(\Sigma_c/2\omega_{\rm rf})| \gg 2\omega_{\rm rf}$,
the $N$-level approximation fails unless it is corrected by including
the intermediate-state Floquet manifold in the model.
For sufficiently
large values of $|\Omega_{\rm c}|$, the probe absorption spectrum in
a 3-level ladder system develops sidebands
induced by the coupling
of the intermediate state to the polarizable upper state, in addition
to the sidebands manifesting in the weak coupling regime. For
certain combinations of the parameters, this spectrum changes
from one exhibiting sharp EIT dips to one exhibiting sharp absorption peaks
as $|\Omega_{\rm c}|$ increases from
 $|\Omega_{\rm c}J_0(\Sigma_c/2\omega_{\rm rf})| \ll 2\omega_{\rm rf}$ to
 $|\Omega_{\rm c}J_0(\Sigma_c/2\omega_{\rm rf})| \gg 2\omega_{\rm rf}$.

%%%%%%%%%%%%%%%%%%%%%%%%%%%%%%%%%%%%%%%%%%%%%%%%%%%%%%%%%%%%%%%%%%%%%%%%%%%%%%%%%%%%%%%%%%%%%%%%%%%%%%%%%%%%%%%%%%%%%%%%%%%%%%%%%%%%%%%%%%%%%%%%%%%%%%%%%%%%%%%%%%%%%%%%%%%%%%%%%%%%%%%%%%%%%%%%%%%%%%%%%%%%%%%%%%%%%%%%%%%%%%%%%%%%%%%%%%%%%%%%%%%%%%%%%%%%%%%%%%%%%%%%%%%%%%%%%%%%%%%%%%%%%%%%%%%%%%%%%%%%%%%%
%%%%%%%%%%%%%%%%%%%%%%%%%%%%%%%%%%%%%%%%%%%%%%%%%%%%%%%%%%%%%%%%%%%%%%%%%%%%%%%%%%%%%%%%%%%%%%%%%%%%%%%%%%%%%%%%%%%%%%%%%%%%%%%%%%%%%%%%%%%%%%%%%%%%%%%%%%%%%%%%%%%%%%%%%%%%%%%%%%%%%%%%%%%%%%%%%%%%%%%%%%%%%%%%%%%%%%%%%%%%%%%%%%%%%%%%%%%%%%%%%%%%%%%%%%%%%%%%%%%%%%%%%%%%%%%%%%%%%%%%%%%%%%%%%%%%%%%%%%%%%%%%
\section*{Acknowledgments}
We thank C.\ S.\ Adams for useful discussions at the early stages of this work
and for his useful comments on the manuscript. We also thank the EPSRC
and the DPST Programme of the Thai Government for financial support.
%%%%%%%%%%%%%%%%%%%%%%%%%%%%%%%%%%%%%%%%%%%%%%%%%%%%%%%%%%%%%%%%%%%%%%%%%%%%%%%%%%%%%%%%%%%%%%%%%%%%%%%%%%%%%%%%%%%%%%%%%%%%%%%%%%%%%%%%%%%%%%%%%%%%%%%%%%%%%%%%%%%%%%%%%%%%%%%%%%%%%%%%%%%%%%%%%%%%%%%%%%%%%%%%%%%%%%%%%%%%%%%%%%%%%%%%%%%%%%%%%%%%%%%%%%%%%%%%%%%%%%%%%%%%%%%%%%%%%%%%%%%%%%%%%%%%%%%%%%%%%%%%
%%%%%%%%%%%%%%%%%%%%%%%%%%%%%%%%%%%%%%%%%%%%%%%%%%%%%%%%%%%%%%%%%%%%%%%%%%%%%%%%%%%%%%%%%%%%%%%%%%%%%%%%%%%%%%%%%%%%%%%%%%%%%%%%%%%%%%%%%%%%%%%%%%%%%%%%%%%%%%%%%%%%%%%%%%%%%%%%%%%%%%%%%%%%%%%%%%%%%%%%%%%%%%%%%%%%%%%%%%%%%%%%%%%%%%%%%%%%%%%%%%%%%%%%%%%%%%%%%%%%%%%%%%%%%%%%%%%%%%%%%%%%%%%%%%%%%%%%%%%%%%%%
%\section*{References}


\begin{thebibliography}{99}
\bibitem{Rabi} Rabi frequencies are assumed to be real, not complex, throughout
the paper.
\bibitem{notesigns} The signs in Eqs.\ (\ref{eq:repb}), (\ref{eq:repb}) and
(\ref{eq:Psiexp}) are chosen in such a way that the eigenstates of 
${\sf H}'_{2{\rm F}}$ belonging to the zero-$\Omega_{\rm c}$ quasienergies
$\epsilon_{b,n}^{(0)}$ and $\epsilon_{c,n}^{(0)}$ defined in
Sec.\ \ref{subsubsec:3a2}
are, respectively,
$|b',n\rangle\rangle$ and $|c',n\rangle\rangle$, not
$|b',-n\rangle\rangle$ and $|c',-n\rangle\rangle$.
\bibitem{notechu} Eqs.\ (\ref{eq:EPERT1}), (\ref{eq:EPERT2}) and 
(\ref{eq:MODEL}) are consistent with the results of Ref.\ \cite{chu}.
\bibitem{footnote2} Perturbation theory gives
${\tilde \Delta}_{\text{c},\pm m} \approx \Sigma_c \pm 2m\omega_{\text{rf}}
\mp {\tilde \delta}_m^{\rm pt}$ with ${\tilde \delta}_m^{\rm pt}=
\sum_{m\neq -n}(\Omega_{\text{c}} J_{m}/2)^2/[2\omega_{\text{rf}}(m+n)]$.
Since ${\tilde \delta}_m^{\rm pt}$ is approximately half the correct shift,
the $N$-level approximation can be improved by replacing
$\Sigma_c-\Delta_{\text{c}}\pm2m\omega_{\text{rf}}$ by
$\Sigma_c-\Delta_{\text{c}}\pm2m\omega_{\text{rf}}\mp{\tilde \delta}_m^{\rm pt}$
in the diagonal elements of the matrix ${\tilde{\sf H}}_{2{\rm F}}^{\prime}$,
or equivalently by replacing
$\Sigma_c-\Delta_{\text{c}}-2k\omega_{\rm rf}$ by
$\Sigma_c-\Delta_{\text{c}}-2k\omega_{\rm rf}+{\tilde \delta}_k^{\rm pt}$
in Eqs.\ (\ref{eq:HRHON1}--\ref{eq:HRHON3}).
\bibitem{newnote} As discussed in Sec.\  \ref {subsubsec:3a2},
the sidebands do not occur exactly at
$\Sigma_c+2n\omega_{\rm rf}$. However, for the parameters of
Fig.\ \ref{fig:Fig7}(a), the difference is too small to be seen.
\end{thebibliography}
\end{document}